\documentclass{aa}  

\usepackage{graphicx}
\usepackage{soul}
\usepackage{txfonts}
\usepackage{xcolor} 
\usepackage{amssymb}
\usepackage{soul} 
\usepackage[normalem]{ulem}
\usepackage{float}
\usepackage{ragged2e}
\usepackage{caption}
\usepackage{subcaption}
\usepackage{rotating}
\usepackage{graphicx}
\usepackage{txfonts}
\usepackage{hyperref}
\usepackage{amsmath}
\usepackage[T1]{fontenc}
\usepackage[utf8]{inputenc}

\def\mpc{{\ \rm Mpc}}
\def\kpc{{\ \rm kpc}}
\def\kms{{\ \rm km\, s^{-1}}}
\usepackage{glossaries}
\glsdisablehyper
\newacronym{sed}{SED}{spectral energy distribution}
\begin{document} 

 \authorrunning{Duplancic et al.}
 \titlerunning{The PAU Survey: AGNs in galaxy pairs}
 \title{The PAU Survey: Active galactic nucleus galaxies in galaxy pairs}

 \author{F. Duplancic\inst{1}\fnmsep\thanks{fduplancic@unsj-cuim.edu.ar},
        M. Siudek\inst{2,3,4},
        D. Navarro-Gironés\inst{5,4,9},
        E. J. Gonzalez\inst{6,7},
        F. Rodriguez\inst{6,8},
        S. Alonso\inst{1},\\
        I. V. Daza-Perilla\inst{10,11},
        D. G. Lambas\inst{6,8},
        E. Gazta\~{n}aga\inst{4,9,12},
        A. L. O'Mill\inst{6,8},
        M. Eriksen\inst{7,13},
        J. Carretero\inst{13,18},\\
        R. Casas\inst{4,9},
        F. J. Castander\inst{4,9},
        J. Garc\'ia-Bellido\inst{14},
        H. Hildebrandt\inst{15},
        M. Mezcua\inst{4,9},
        R. Miquel\inst{7,16},\\
        C. Padilla\inst{7},
        P. Renard\inst{4,9,17},
        E. S\'anchez\inst{18},
        S. Serrano\inst{4,9,19},
        I. Sevilla-Noarbe\inst{18},
        P. Tallada-Cresp\'i\inst{13,18}
        }

 \institute{
        $^{1}$Gabinete de Astronomía Extragaláctica, Departamento de Geofísica, Astronomía y Física, CONICET, Facultad de Ciencias Exactas, Físicas y Naturales, Universidad Nacional de San Juan, Av. Ignacio de la Roza 590 (O), J5402DCS, Rivadavia, San Juan, Argentina.\\
        $^{2}$Instituto de Astrof\'{\i}sica de Canarias, V\'{\i}a L\'actea, 38205 La Laguna, Tenerife, Spain.\\
        $^{3}$Instituto de Astrof\'isica de Canarias (IAC); Departamento de Astrof\'isica, Universidad de La Laguna (ULL), 38200, La Laguna, Tenerife, Spain.\\
        $^{4}$Institute of Space Sciences (ICE, CSIC), Campus UAB, Carrer de Can Magrans, s/n, 08193 Barcelona, Spain.\\
        $^{5}$Leiden Observatory, Leiden University, Einsteinweg 55, 2333 CC Leiden, the Netherlands.\\
        $^{6}$CONICET. Instituto de Astronomía Teórica y Experimental (IATE), Laprida 854, Córdoba X5000BGR, Argentina.\\
        $^{7}$Institut de Física d’Altes Energies (IFAE), The Barcelona Institute of Science and Technology, Campus UAB, 08193 Bellaterra (Barcelona), Spain.\\
        $^{8}$Universidad Nacional de Córdoba (UNC). Observatorio Astronómico de Córdoba (OAC), Laprida 854, Córdoba X5000BGR, Argentina.\\
        $^{9}$Institut d'Estudis Espacials de Catalunya (IEEC), E-08860 Castelldefels (Barcelona), Spain.\\
        $^{10}$Center for Space Science and Technology, University of Maryland, Baltimore County, 1000 Hilltop Circle, Baltimore MD 21250, USA.\\
        $^{11}$Center for Research and Exploration in Space Science and Technology, NASA/Goddard Space Flight Center, Greenbelt, MD 20771, USA.\\
        $^{12}$Institute of Cosmology \& Gravitation, University of Portsmouth, Dennis Sciama Building, Burnaby Road, Portsmouth PO1 3FX, UK.\\
        $^{13}$Port d'Informaci\'o Cient\'ifica (PIC), Campus UAB, C. Albareda s/n, 08193 Bellaterra (Barcelona), Spain.\\
        $^{14}$Instituto de Física Teórica CSIC/UAM, Universidad Autónoma de Madrid, 28049 Madrid, Spain.\\
        $^{15}$Ruhr University Bochum, Faculty of Physics and Astronomy, Astronomical Institute (AIRUB), German Centre for Cosmological Lensing, 44780 Bochum, Germany.\\
        $^{16}$Institució Catalana de Recerca i Estudis Avançats (ICREA), 08010 Barcelona, Spain.\\
        $^{17}$Department of Astronomy, Tsinghua University, Beijing 100084, China.\\
        $^{18}$Centro de Investigaciones Energéticas, Medioambientales y Tecnológicas (CIEMAT), Avenida Complutense 40, 28040 Madrid, Spain.\\
        $^{19}$Satlantis, University Science Park, Sede Bld 48940, Leioa-Bilbao, Spain.
      }
    \date{Received xx; accepted xx}

 
  \abstract
   {}
   {Leveraging the high-precision photometric redshifts of the Physics of the Accelerating Universe Survey (PAUS), we aim to investigate the connection between active galactic nuclei (AGNs) and galaxy pairs up to redshifts of $\sim1.0$.}
   {We identified 16,824 galaxy pairs with projected separations of $r_p < 100$ kpc and radial-velocity differences of $\Delta V < 3500$ km s$^{-1}$. AGNs were selected by cross-matching PAUS galaxies with multi-wavelength AGN catalogues, yielding 270 AGNs in pairs out of a parent sample of 8,880 AGNs in PAUS. We constructed control samples of isolated AGNs; non-AGNs in pairs; and isolated non-AGNs matched in redshift, absolute magnitude, and environment.   }
    { Active galactic nuclei in pairs show progressively redder rest-frame colours with increasing redshift, particularly at redshifts of $z\gtrsim 0.5$, whereas isolated AGNs and non-AGNs in pairs show no significant colour evolution and remain at $\mathrm{M_g-M_r}\sim0.65$. Pairs hosting AGNs are also preferentially morphologically disturbed interacting systems found at smaller projected separations and velocity offsets, consistent with more dynamically advanced interactions.}  
   {These results provide evidence of a redshift-dependent link between galaxy interactions and AGN activity, suggesting that interactions play an increasingly important role in triggering nuclear activity at intermediate redshifts. This study highlights the unique capability of PAUS to probe AGN--galaxy interaction connections beyond the local Universe with unprecedented statistical power.}


\maketitle

\section{Introduction}
Active galactic nuclei (AGNs) represent a key phase in galaxy evolution, providing insights into the co-evolution of supermassive black holes (SMBHs) and their host galaxies. Galaxy interactions have long been proposed as effective triggers for AGN activity, as tidal forces during close encounters can drive gas inflows towards central regions, potentially feeding SMBHs \citep[e.g.][]{DiMatteo2005, Hopkins2008}. Understanding when and how these interactions trigger nuclear activity across cosmic time is essential for constraining galaxy-evolution models \citep{AlexanderHickox2012,Harrison2024}.

In the local Universe, several studies point to a connection between galaxy interactions and AGN triggering. Observational studies of galaxy pairs show enhanced AGN fractions compared to isolated galaxies \citep{Alonso2007,Koss2011,Ellison2013,Silverman2011}. This enhancement becomes more pronounced as pair separation decreases and extends through the merger sequence, with peak AGN activity often observed in post-merger systems \citep{Ellison2013,Goulding2018}. Multi-wavelength studies further strengthen this evidence; for example, \citet{Li2023} identified AGNs using multi-wavelength diagnostics (X-ray, infrared, radio, and optical), finding a 36\% excess of AGNs in post-mergers compared to non-interacting control samples. 

In addition, different observational studies have established a connection between galaxy morphological disturbances, such as tidal features and bridges, and nuclear activity \citep[e.g.][]{Koss2010,Ellison2011,Mesa2014,Hernandez-Toledo2023}. These results support the scenario in which gravitational interactions and mergers can trigger gas inflows towards the central regions, feeding the SMBHs and enhancing the activity of AGNs, particularly in galaxies displaying clear morphological signatures of interaction.

The role of obscuration also appears to be relevant in the identification of AGNs, with mid-infrared (MIR) selected AGNs becoming increasingly prevalent, compared to optical AGNs, as pair separation decreases \citep{Satyapal2014}. For instance, merging systems are 5--17 times more likely to host dust-obscured AGNs than non-interacting galaxies \citep{Weston2017,Secrest2020}. These findings suggest that interactions may preferentially trigger dust-obscured nuclear activity, potentially contributing to discrepancies between AGN-selection methods.

Environmental factors introduce additional complexity. The local galaxy density affects the activity of AGNs independently of pair interactions \citep[e.g.][]{Kauffmann2003,Sabater2013,Coldwell2017}, with evidence suggesting that AGN fractions peak in small-scale
associations rather than in larger groups \citep{Duplancic2021}. In addition, large-scale environments can influence the activity of AGNs, with studies showing different perspectives. \citet{Koulouridis2024} report fewer active galaxies in cluster centres compared to their outskirts, implying environment-dependent triggering mechanisms. In contrast, \citet{Rembold2024} find minimal differences in group mass and density between AGNs and control galaxies, suggesting limited environmental impact relative to galaxy interactions. Meanwhile, \citet{Ceccarelli2022} highlighted voids as favourable for AGN activity, proposing that secular galaxy evolution and gas inflows from the void interior can fuel central black holes, thus supporting AGN activity. Observations of AGNs in dwarf galaxies further indicate that while mergers can trigger nuclear activity, secular processes are often the dominant mechanism in these low-mass hosts \citep{Erostegui2025}. However, this well-established paradigm faces significant challenges when extended beyond the local Universe. Recent studies using deeper surveys ($0.3<z<2.5$) have failed to detect a statistically significant excess of AGNs in pairs and post-merger systems or dense environments \citep{Shah2020,Silva2021,Siudek2023,Dougherty2024}. 

These discrepancies in different studies raise some questions, including whether the interaction--AGN connection evolves with redshift or if current observational limitations obscure the relationship.
Addressing these questions requires the precise identification of galaxy pairs across a wide redshift range, a task that has proven challenging with conventional techniques. Spectroscopic surveys provide accurate redshifts but suffer from limited area coverage and fibre collision constraints that systematically exclude close pairs. Traditional broad-band (BB) photometric surveys cover larger areas but lack the redshift precision necessary to reliably distinguish physical pairs from chance projections, particularly at intermediate redshifts.

In this context, the Physics of the Accelerating Universe Survey \citep[PAUS;][]{Benitez2009} offers significant advances to overcome these limitations by combining wide-area coverage with high-precision photometric redshifts (photo-z). Using 40 narrow-band (NB) filters, PAUS achieves a redshift precision of $\sigma_{68} \leq 0.02$ for galaxies with $i_{AB} < 23$ in the z=0.10-1.0 redshift range, approaching the accuracy of low-resolution spectroscopy \citep[][]{Eriksen_2020,navarro_photoz,Daza-Perilla2025}. This level of precision enables the reliable identification and characterisation of galaxy pairs at intermediate redshifts ($0.5<z<1.0$) as presented in \citet{rodriguez2020close} and \citet{gonzalez2023pau}. These authors also analysed galaxy pair properties, such as mean mass and luminosity. Moreover, the analysis of the galaxy pairs identified included an estimated relation between the mean halo mass and luminosity of these systems, taking advantage of the lensing properties available in the PAUS fields.

The $z\sim0.5-1.0$ redshift range represents an important epoch in the study of the interaction-AGNs connection. This period falls after the peak of cosmic star formation and AGN activity \citep[$z\sim1.5-3.0$;][]{Madau2014,Aird2015}, offering insights into how AGNs' triggering mechanisms evolve from cosmic noon to the present day. This transition zone --- where local Universe studies show strong interaction--AGN correlations but higher redshift investigations find minimal connection --- offers unique insights into the redshift and time evolution of AGN triggering mechanisms.

In this study, we exploited the high-precision photo-z provided by PAUS to investigate the connection between AGNs and galaxy pairs up to redshift $z\sim1.0$.
This paper is organised as follows. Section \ref{data} presents the PAUS dataset and photo-z used in this study. In Sect. \ref{met} we describe the methodology used to select galaxy pairs, the identification of AGNs in PAUS data, and the construction of suitable control samples. 
Section \ref{results} presents the results of the study of AGNs in pairs and the comparison with suitable control samples. Finally, Sect. \ref{conc} summarises our main conclusions. Throughout this work, we adopted a cosmological $\Lambda$CDM model characterised by $\Omega_m=0.25$, $\Omega_{\Lambda}=0.75$ and $H_0=69~h~\kms \mpc $ parameters.

\section{Data}
\label{data}

\subsection{PAUS}
PAUS is a photometric survey conducted using the William Herschel Telescope at El Roque de Los Muchachos observatory in the Canary Islands \citep{Benitez2009,Marti2014,Serrano2023}. The PAUCamera \citep{Padilla2019} used a set of 40 NB filters ranging from 4500 to 8500\AA\ in steps of 100\AA\, with a full width at half maximum (FWHM) of 130\AA. This exceptional filter arrangement is designed to provide high-precision photo-z, outperforming the precision achievable with BB observations \citep{Eriksen_bcnz,Eriksen_2020, Alarcon2021,Soo2021,Cabayol2023,navarro_photoz,Daza-Perilla2025}.  

PAUS targets include the COSMOS field \citep{Scoville2007}, used mainly for calibration and validation; the W1, W3, and W4 Wide Fields from CFHTLenS \citep{Heymans2012,Erben2013}; and the GAMA G09 field, which overlaps with KiDS \citep{deJong2013}. PAUS has performed only limited observations in the W4 field; therefore, the W1, W3, and G09 fields compose the PAUS deep wide fields. The positions and areas of these fields are summarised in Table \ref{PAUSfields}. The survey covers an area of $\sim$43 deg$^2$ in all 40 NB filters and $\sim$51 deg$^2$ with at least 30 NB filters, reaching $i_{AB}=23$ with $\sim$1.8 million objects. For a detailed explanation of the photometry and flux calibration, we refer the reader to \citet{Serrano2023} and \citet{Castander2024}.

In this work, the G09 field was excluded because local overdensity measurements -- a critical parameter for generating our control samples -- were unavailable for this area. It is also worth noting that W1 and W3 are remarkably consistent in terms of depth, with a number-density difference between them of less than 3\% \citep{navarro_photoz}. To achieve our objectives, we merged the W1 and W3 fields into a single PAUS master catalogue, providing a homogeneous dataset with consistent magnitude depth and observational parameters for the subsequent analysis.

\begin{table}[ht]
\centering
\caption{Area and RA/Dec limits of the PAUS fields.}
\begin{tabular}{lccc}
\hline\hline
Field & Area & RA limits & Dec limits \\
      & (deg$^2$) & (deg) & (deg) \\
\hline
W1    & 12.04 & [32.3, 38.5]  & [-6.1, -4.0] \\
G09* & 15.70 & [131.0, 139.0] & [-1.7, 0.5] \\
W3    & 22.64 & [209.0, 219.5] & [51.4, 55.6] \\
\hline
\multicolumn{4}{l}{\footnotesize *Field excluded from the final analysis of this work.}
\end{tabular}
\label{PAUSfields}
\end{table}

\subsection{Photometric redshifts}

For $\sim1.5$ million PAUS galaxies there are two different photometric redshift catalogues, {\sc BCNZ2} \citep{navarro_photoz} and {\sc DEEPz} \citep{Daza-Perilla2025}. The main difference between these methods is that BCNZ2 is a classic approach based on spectral-energy-distribution (SED) template-fitting methods, while DEEPz is a deep-learning-based photo-z method that utilises neural networks.
We note that the {\sc BCNZ2} templates do not include AGN components, which could, in principle, affect photometric redshift estimates for AGNs. In contrast, {\sc DEEPz} was trained on datasets that include a small fraction of AGNs ($\sim$5\% in W1+W3 and $\sim$3\% in G09+COSMOS). In Appendix \ref{sec:photo-z_accuracy} we describe how we used well-controlled samples of AGNs and non-AGNs  to test the photometric redshift performance of {\sc BCNZ2} and {\sc DEEPz} redshifts. In both cases, we do not find significant differences in photometric redshift accuracy (characterised by the $\sigma_{\mathrm 68}$, the outlier rate and the bias) between spectroscopically confirmed AGNs and inactive galaxies. We therefore conclude that AGN-related biases in the photo-z used in this work are negligible. 
However, we observe that {\sc DEEPz} achieves better precision for faint galaxies ($i > 20$), with improvements of up to $\sim$40\% compared to {\sc BCNZ2}. Despite this, the redshift distribution produced by {\sc DEEPz} exhibits discrete features (see Fig.~\ref{zdists}) that can introduce artificial clustering at specific redshift intervals. This behaviour significantly affects the identification of close galaxy pairs, particularly those with small redshift separations.

For this reason, and to ensure consistency with previous PAUS studies of galaxy pairs \citep{rodriguez2020close,gonzalez2023pau}, we adopt the weighted version of the {\sc BCNZ2} photo-z estimate (hereafter $z_\mathrm b$). This choice provides reliable pair identification while maintaining compatibility with existing pair catalogues.

The final photo-z catalogue spans the $0 < z_b \leq 2$ range. Rest-frame absolute magnitudes and colours were computed from the observed PAUS photometry using photo-z and $k$-corrections derived from the best-fitting SED templates, assuming a standard $\Lambda$CDM cosmology. Observed magnitudes were corrected for Galactic extinction prior to the calculation. Stellar masses were taken from \citet{Csizi2024}, whose authors derived galaxy physical properties through SED-fitting techniques applied to the PAUS-COSMOS sample. These quantities were used consistently throughout this work to characterise the galaxy populations.

\section{Methodology}
\label{met}
Our goal is to investigate the impact of galaxy interactions on nuclear activity by analysing the occurrence of AGNs in galaxy pairs. To this end, we constructed well-defined samples of galaxy pairs, as well as control samples of isolated galaxies in which we identify active objects. These samples must exhibit similar distributions in redshift, absolute magnitude (a proxy of stellar mass), and environment \citep{Perez2009}. In the following sections, we describe the methodology used to build the samples used in this study.

\subsection{Pair catalogue}
\label{paircat}

To detect pairs using photo-z, we followed the procedure developed and implemented in \cite{rodriguez2020close}, \cite{gonzalez2023pau}, and \cite{cerdosino2024structure}. These authors identified galaxy pairs by analysing their angular and redshift proximity while accounting for the photo-z error inherent to PAUS. In addition, the identification algorithm applies an isolation criterion designed to exclude pairs embedded in galaxy groups or clusters.

We define a pair as two galaxies fulfilling the following selection criteria:
\begin{itemize}
    \item Projected distance ($r_p$): less than 100 $\kpc$. It was computed from the angular distance between the two galaxies and the co-moving distance to the central galaxy,
    \[
    r_p = D_{\rm c}(z_{\rm c}) \, \tan(\theta),
    \]  
where $\theta$ is their angular separation on the sky, $z_{\rm c}$ is the redshift of the central galaxy, and $D_{\rm c}(z_{\rm c})$ is the co-moving distance to the central galaxy evaluated in the adopted cosmology. This limit allowed us to probe a range of interaction stages between physically associated systems.

    \item Velocity difference ($\Delta V$): less than 3500$\kms$. It was estimated from the redshift separation as  
    \[
    \Delta V = c\,|\Delta z|,
    \]  
where $\Delta z = z - z_{\rm c}$ is the redshift difference between a given galaxy (with redshift $z$), and the central galaxy (with redshift $z_{\rm c}$), and $c$ is the speed of light. The adopted threshold, motivated by simulations \citep{rodriguez2020close}, yields an optimal balance between purity and completeness given the precision of the photo-z.

    \item Absolute magnitude: at least one galaxy must have an absolute magnitude in the $r$ band brighter than $-17.5$. This ensures $r$-band completeness across the redshift range of the sample.
    \item Magnitude difference: the magnitude difference between the two galaxies must not exceed $2.5$. This criterion was adopted to minimise the contamination from bright central galaxies with faint satellites.
    \item Isolation criterion: within a radius of 300$\kpc$, there must be no neighbouring galaxy with a magnitude difference relative to the central galaxy smaller than $2.5$, nor with a velocity difference less than $3500\kms$. This guarantees the pair is the dominant system in its local environment, reducing false positives from richer structures.
\end{itemize}

Using these criteria, we identified a total of 11\,635 galaxy pairs in the PAUS master catalogue\footnote{3847 in W1,  and 7788 in W3}. The level of contamination in the photometrically selected pair catalogue was evaluated through an independent analysis of a subsample of pairs with spectroscopic redshifts for both members (spec--spec pairs), presented in Appendix~\ref{Pairstest}. This analysis provides an upper limit on the fraction of chance projections.

\begin{figure}[ht]
\centering
        \includegraphics[width=.40\textwidth]{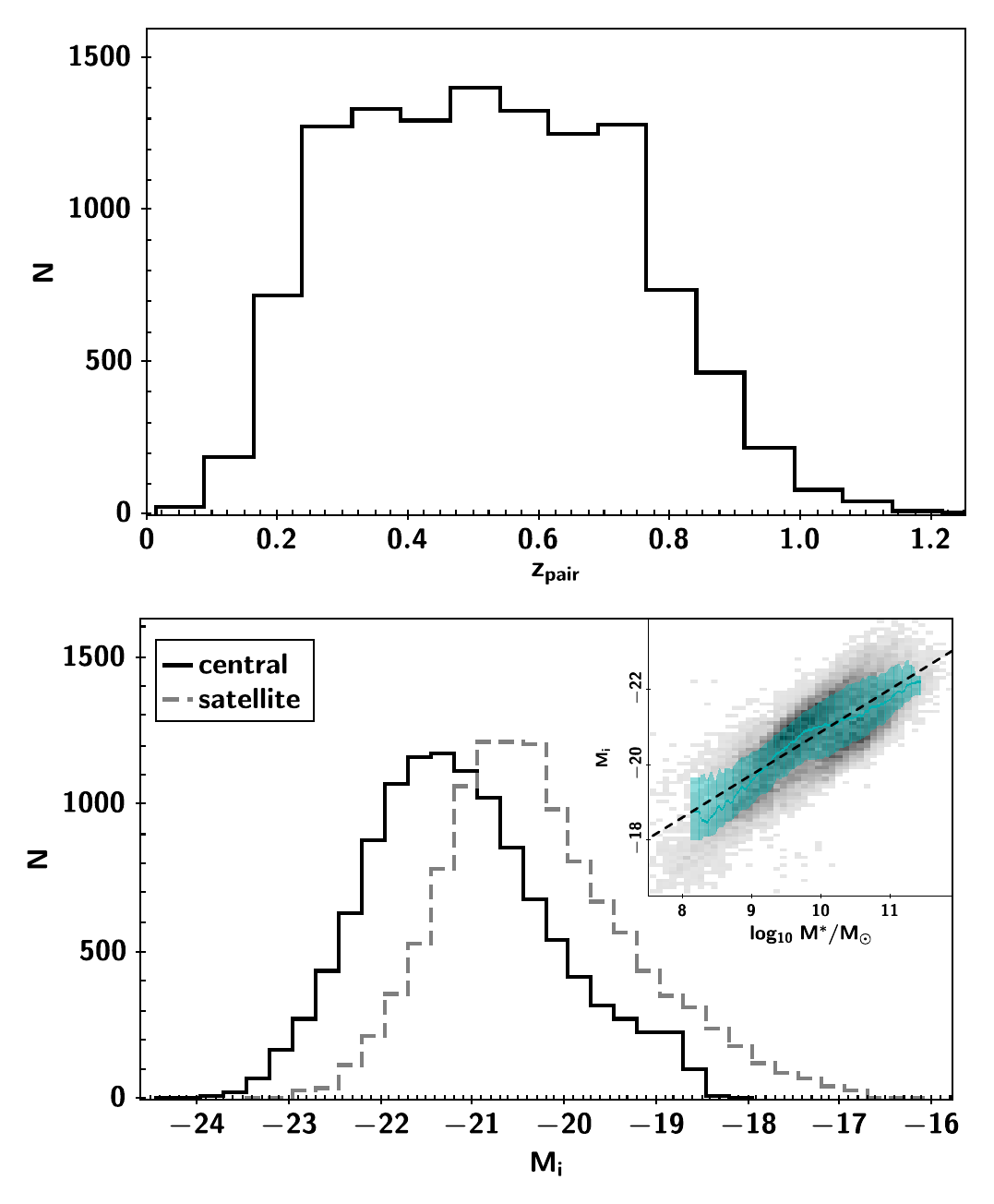}
    \caption{Distribution of pair photo-z ($z_\mathrm {pairs}$; \textit{top}), $i$-band absolute magnitude ($M_i$) of the brightest and the faintest galaxy in the system (\textit{bottom}). The inset shows the relation between $M_i$ and stellar mass ($\rm log(M{*}/M_{\odot})$) for galaxies in the pair sample.
    }
    \label{pairzMi}
\end{figure}

The photo-z pair distribution (calculated by averaging the photo-z of the central (brightest) and satellite (faintest) galaxy in the system) and the i-band absolute magnitude ($M_i$) distribution of the central and satellite galaxies in the pair sample are shown in Fig. \ref{pairzMi}. It is clear that pairs can be adequately identified up to $z_{\mathrm{pair}}\sim 1.2$ and that, on average, the i-band absolute-magnitude difference between the brightest and the faintest galaxy in the system is $\sim 1$ magnitude.
In this plot we also show the relation between $M_i$ and stellar masses ($\rm log(M{*}/M_{\odot})$) for pair galaxies,
finding a suitable correlation between both parameters. 
Given that pair selection is based on photometric criteria, we consider the  $\mu=10^{-0.4(M_\mathrm{is}-M_\mathrm{ic})}$ parameter as a photometric proxy of the mass ratio;\footnote{This parameter ranges from 0 to 1 and quantifies the similarity in stellar mass between the two galaxies.} here, $M_\mathrm{ic}$ and $M_\mathrm{is}$ are the i-band absolute magnitude of the brightest (central) and the faintest (satellite) galaxy in the pair, respectively. 
We found that 70\% of the pairs in our sample have $\mu>1/3$ and can therefore be considered as major mergers according to the definition of \citet{Hopkins2010,Lambas2012, RodriguezGomez2015}.

\subsection{AGN identification}
\label{AGNsec}

To identify AGNs in PAUS, we used a 1-arcsec matching radius to cross-match with external AGN catalogues. This choice reflects the sub-arcsecond astrometric accuracy of the CFHTLS and KiDS reference images used for PAUS photometry. We rely on a set of multi-wavelength AGN catalogues and diagnostics, providing complementary views of nuclear activity.

For mid-infrared selection (MIR-AGNs), we used the R90 catalogue from \citet{Assef2018}, which includes $\sim$4.5 million active galaxy candidates selected from the Wide-field Infrared Survey Explorer AllWISE data release \citep[][]{Cutri2013}. The AGN selection is based on colour-colour cuts in the ($W2 - W3$) versus ($W1 - W2$) plane, where $W1$, $W2$, and $W3$ correspond to the WISE bands at 3.4, 4.6, and 12~$\mu$m, respectively. The 'R90’ designation refers to a 90\% reliability threshold, meaning that at least 90\% of the sources classified as AGNs are expected to be genuine.
This MIR-based approach is highly efficient at identifying dust-obscured AGNs, since ultraviolet (UV) and optical emission from the accretion disc absorbed by circumnuclear dust is reprocessed and re-emitted at MIR wavelengths.

Regarding optical emission-line diagnostics (BPT-AGNs), we used the SDSS Portsmouth emission-line catalogue \citep[\texttt{emissionLinesPort}; see][]{Thomas2013}, which was queried via SQL through the SDSS Catalogue Archive Server Jobs System (CasJobs\footnote{http://skyserver.sdss.org/CasJobs/}). This table provides a classification \texttt{bpt} entry derived from BPT diagrams \citep{Baldwin1981} considering the classification scheme of \citet{Kauffmann2003}, \citet{Kewley2001}, and \citet{Schawinski2007}, applied to galaxies with H$\alpha$, H$\beta$, [O\,\textsc{iii}]$\lambda5007$, and [N\,\textsc{ii}]$\lambda6584$ measurable emission lines, with an amplitude-to-noise ratio of A/N $>$ 1.5.  The BPT diagram separates galaxies according to the dominant ionisation mechanism that produces their optical emission lines. In AGNs, the intense radiation field from accretion onto SMBHs photoionises the surrounding gas, yielding higher excitation and stronger forbidden lines such as [O\,\textsc{iii}], allowing the distinction of AGNs from star-forming galaxies. This optical method identifies unobscured or moderately obscured AGNs with prominent narrow emission lines.   
Possible \texttt{bpt} values are `BLANK' (if emission lines are unavailable), `Star Forming', `Seyfert', `LINER', `Seyfert/LINER', or `Composite'. We selected AGNs classified as Seyfert, LINER, or Seyfert/LINER. 

For X-ray selection (Xray-AGNs), we used the probabilistic classification of \citet{Tranin2022}, which combines X-ray data from the 2SXPS \citep{Evans2020} and 4XMM-DR10 \citep{Webb2020} catalogues. The authors applied a Bayesian approach to assign to each source the probability of belonging to different astrophysical classes (e.g. AGN, star, X-ray binary, cataclysmic variable) using X-ray spectral and temporal properties together with multi-wavelength information. The resulting catalogue includes three classification parameters: \texttt{class}, which gives the most probable source type; \texttt{classification}, which provides the confidence level of the association; and \texttt{alternative}, which lists secondary possible identifications when ambiguity remains. In this work, we considered sources labelled 'active galaxy' in any of these parameters as X-ray AGNs. X-ray selection is one of the most direct and efficient methods for identifying AGNs, since the high-energy photons originate in the innermost regions of the accretion disk and corona surrounding the SMBHs. In these environments, Compton up-scattering of UV photons by hot electrons produces a power-law X-ray spectrum that is largely unaffected by stellar processes, making X-ray emission a reliable AGN tracer even in the presence of dust obscuration.  
    
Finally, for the quasar census (MQS-AGNs) we used the Million Quasars (Milliquas) Version 8 catalogue, compiled by \citet{Flesch2023}. It is a comprehensive database of quasars, AGNs, and photometric AGN candidates. Combining spectroscopic and photometric identifications, Milliquas provides coverage of a broad redshift range, including faint and high-redshift AGNs. This catalogue is based on literature sources \citep[see,][]{Flesch2023} and comprises 907\,144 Type I quasars and AGNs, together with 2814 blazars, 45\,816 Type II objects, and 66\,026 radio/X-ray-associated quasar candidates with $p_{\mathrm {quasar}} = 99\%$; it is largely complete based on literature sources. This catalogue extends our AGN census to include luminous unobscured quasars and high-redshift active galaxies, providing a complementary view of nuclear activity across cosmic time.

\begin{figure}[ht]
\centering
        \includegraphics[width=.49\textwidth]{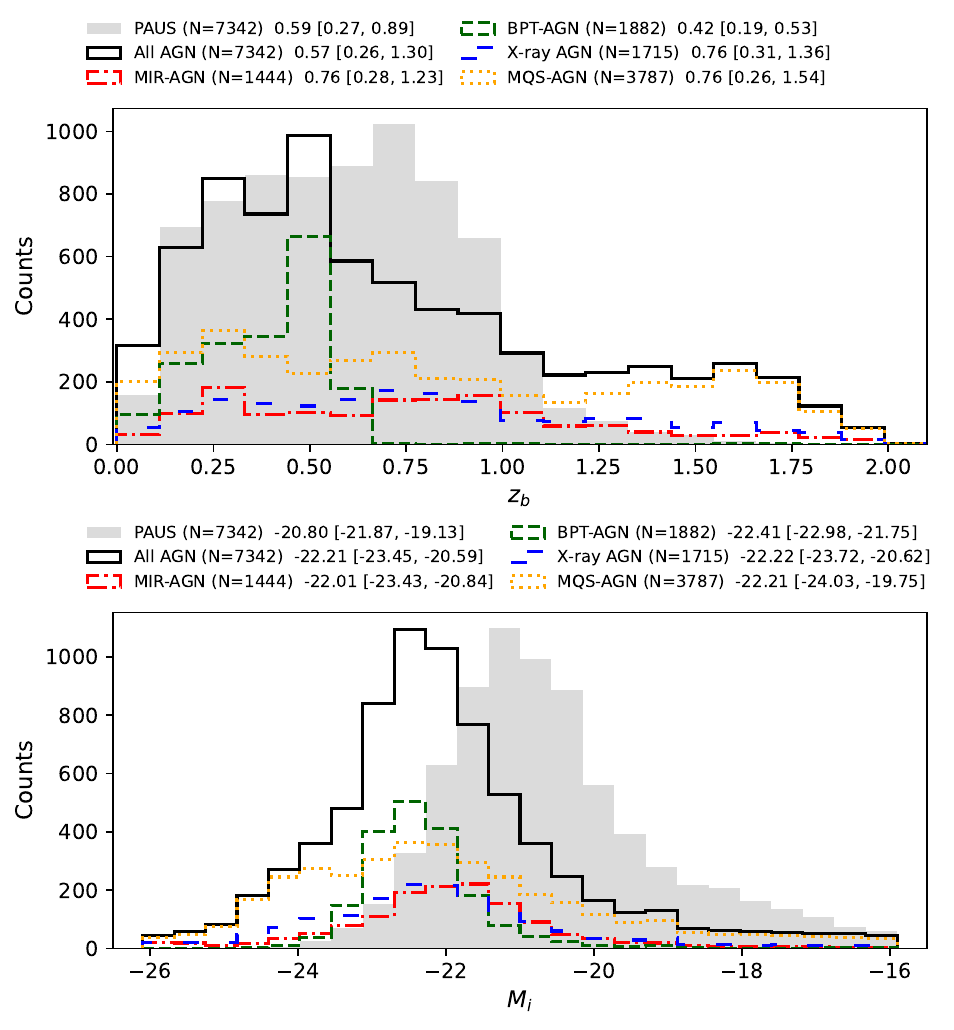}
    \caption{Distributions of photo-z ($z_\mathrm b$; \textit{top}) and $i$-band absolute magnitude ($M_i$; \textit{bottom}) for the sample of 7342 AGNs and for each AGN subsample. We also included the distribution of 7342 non-AGN PAUS galaxies. At the top of the panels we show the number of galaxies, median value, and $Q_1$ and $Q_3$ percentiles of each sample.
    }
    \label{z_Mi_dist_AGNs}
\end{figure}

 It is worth noting that PAUS fields are fully covered by the different AGN catalogues.
Integrating samples across MIR, X-ray, optical, and quasar catalogues provides a comprehensive perspective on the AGN population. This multi-wavelength analysis enhances the reliability of AGN identification and may also help to explore AGN evolution and their role in large-scale cosmic structures. 
In Appendix \ref{testing_AGNs} we study the overlap and intersection of the different AGN classification methods used in this work. This analysis reveals the consistency and complementarity of the different AGN selection techniques within our galaxy sample.

We found 7\,342 AGNs in the PAUS master catalogue. In Fig. \ref{z_Mi_dist_AGNs} we plot $i$-band absolute magnitude ($M_i$)  and photo-z ($z_\mathrm b$) distributions for the different AGN categories identified in the PAUS data. We also included the distribution of the total AGN sample and a random sample of  7342 PAUS galaxies not identified as AGNs. In the figure we list the median values of $M_i$ and $z_\mathrm b$ along with the 25th and 75th percentiles ($Q_1$ and $Q_3$) of each distribution. 

From Fig. \ref{z_Mi_dist_AGNs} we observe different behaviours for AGNs compared with non-AGN PAUS galaxies. In all cases, the Kolmogorov--Smirnov statistic between the different AGN samples and PAUS regular galaxies gives extremely small p values, much smaller than the common $\alpha=0.05$ threshold. This implies, in all comparisons, a rejection of the null hypothesis that the samples are drawn from the same distribution. Also, this figure shows that the distributions of all AGN subsamples are brighter than PAUS galaxies and have a larger median redshift, except for BPT-AGNs located closer than $z=0.65$. This is because the $\mathrm H_\alpha$ emission line is not detectable at optical wavelengths beyond this redshift. The MIR- and X-ray-selected AGNs show the highest median redshifts, reflecting their sensitivity to more distant and dust-obscured systems. The MQS-AGNs sample shows the widest distributions in both absolute magnitude and photo-z, indicating a heterogeneous population spanning a broad range of luminosities and cosmic epochs.
These results highlight the necessity of building suitable control samples to compare the properties of non-AGN PAUS galaxies with those of AGNs.

\subsection{Control samples}
\label{control_samples}

Building robust comparison samples is essential for accurately assessing the impact of interactions on galaxy properties and evolution. We used PAUS identification numbers to cross-match galaxies in the pair catalogue with the PAUS AGN sample and identified  214 (91 in W1 and 123 in W3) active galaxies in pairs (including only two dual-AGN systems). To evaluate the effect of the pair environment on nuclear activity, we also constructed a sample of isolated galaxies using a modified version of the pair-finding algorithm, selecting single galaxies that satisfied the same isolation criteria as pairs. This isolated sample comprises 832\,248 galaxies and provides a baseline for analysing galaxy properties in the absence of close companions.

Although the isolation criterion prevents galaxies from being embedded in rich structures on small scales, it does not capture variations in the large-scale environment. To control for these effects, we employed the luminosity density contrast, $\delta$, as an additional parameter characterising the background density. This quantity measures local fluctuations in the luminosity density relative to the cosmic mean, assuming that luminosity traces the underlying mass distribution. We computed $\delta$ from \texttt{HEALPix} maps constructed in co-moving shells of 5~Mpc$/h$, where each pixel stores the integrated luminosity of galaxies within its projected area. The local luminosity density, $\rho_\text{cylinder}$, was estimated within a cylindrical volume of 15~Mpc$/h$ length and 7.5~Mpc$/h$ radius, summing the luminosities of intersecting pixels. The luminosity contrast density is obtained as
\begin{equation}
    \delta = \frac{\rho_\text{cylinder} - \rho_\text{shell}}{\rho_\text{shell}},
\end{equation}
where $\rho_\text{shell}$ is the total luminosity density in the co-moving shell. Each galaxy was assigned a $\delta$ value based on the nearest pixel, allowing the construction of control samples with matched large-scale environments and minimising biases from cosmic structure. It is worth noting that appropriate masks were applied to account for the geometry of the PAUS fields. The density maps derived with this method are also being used to study galaxy clustering in PAUS data (Gonzalez et al., in prep).

\begin{figure}[!h]
\centering
        \includegraphics[width=.45\textwidth]{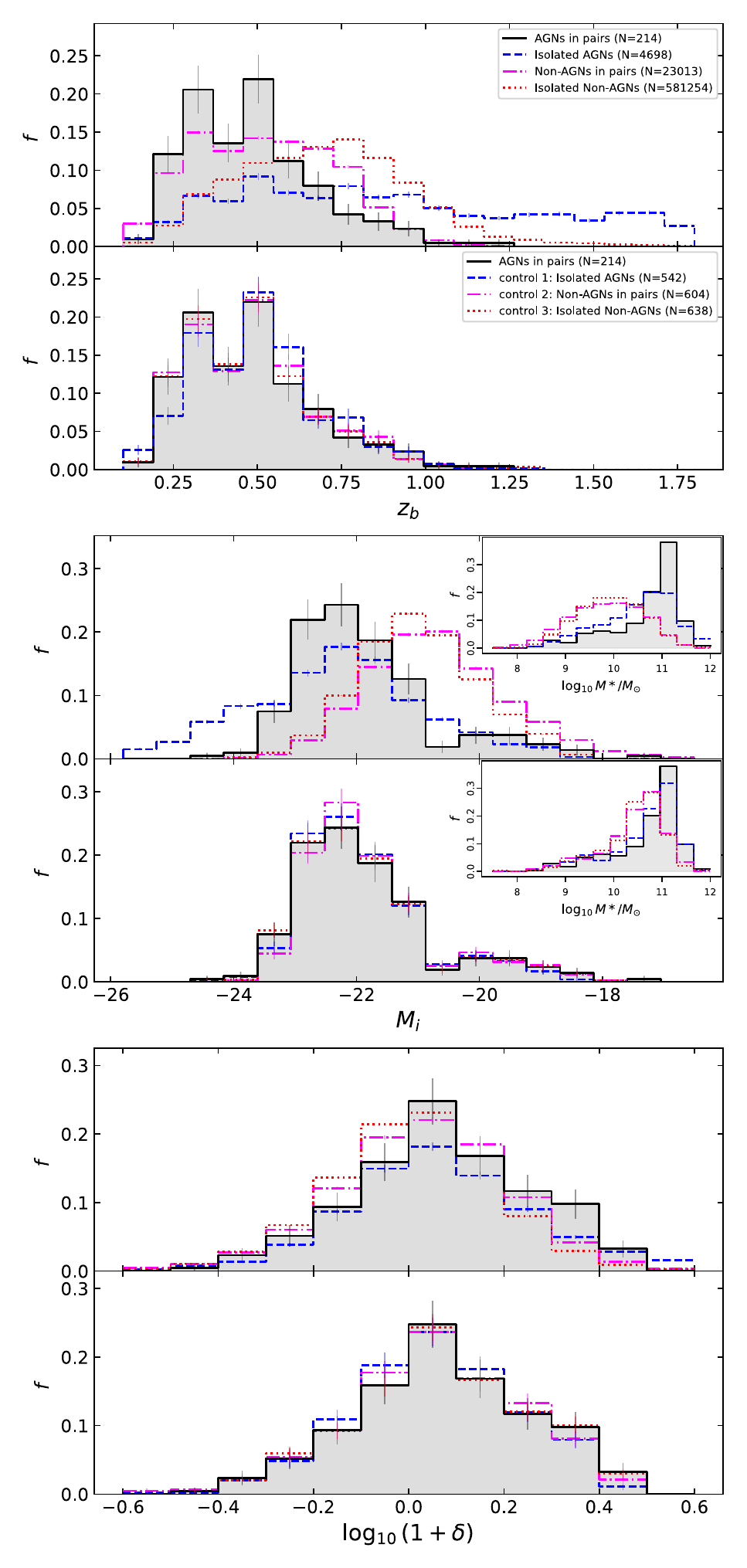}
    \caption{Normalised distributions of redshift ($z_\mathrm{b}$; \textit{top}), absolute $i$-band magnitude ($M_i$; \textit{middle}), and overdensity ($\delta$; \textit{bottom}). For each parameter, the upper panels show the distributions of the original samples: AGNs in pairs (black line with grey shading), isolated AGNs (dashed blue line), non-AGNs in pairs (dot-dashed~magenta line), and isolated non-AGNs (dotted red line). The lower panels display the corresponding distributions for the control samples matched to the AGNs in pairs: \textsc{control 1} (isolated AGNs, dashed blue line), \textsc{control 2} (non-AGNs in pairs, dot-dashed magenta line), and \textsc{control 3} (isolated non-AGNs, dotted red line). The insets in the $M_i$ panels show the resulting stellar mass distributions ($\log_{10} M_{\star}/M_{\odot}$) for the original (top) and matched (bottom) samples. Error bars represent Poissonian uncertainties for each bin.}
    \label{cont1}
\end{figure}

We used the galaxy-pair and isolated-galaxy catalogues to define four samples: AGNs in pairs, which constitute our reference sample; isolated AGNs (\textsc{control 1}); non-AGNs in pairs (\textsc{control 2}); and isolated non-AGNs (\textsc{control 3}). The upper panels of Fig.~\ref{cont1} show the distributions of redshift, absolute $i$-band magnitude, and local overdensity for these original samples, together with their respective number counts. These distributions differ significantly from those of the reference AGNs in pairs, highlighting the need to construct control samples that reproduce the intrinsic properties of the reference population. Control samples were therefore built to match the distributions of redshift, absolute $i$-band magnitude, and overdensity of AGNs in pairs, adopting maximum allowed differences of 0.05, 0.25, and 0.1, respectively.

\begin{figure}[h]
\centering
        \includegraphics[width=.48\textwidth]{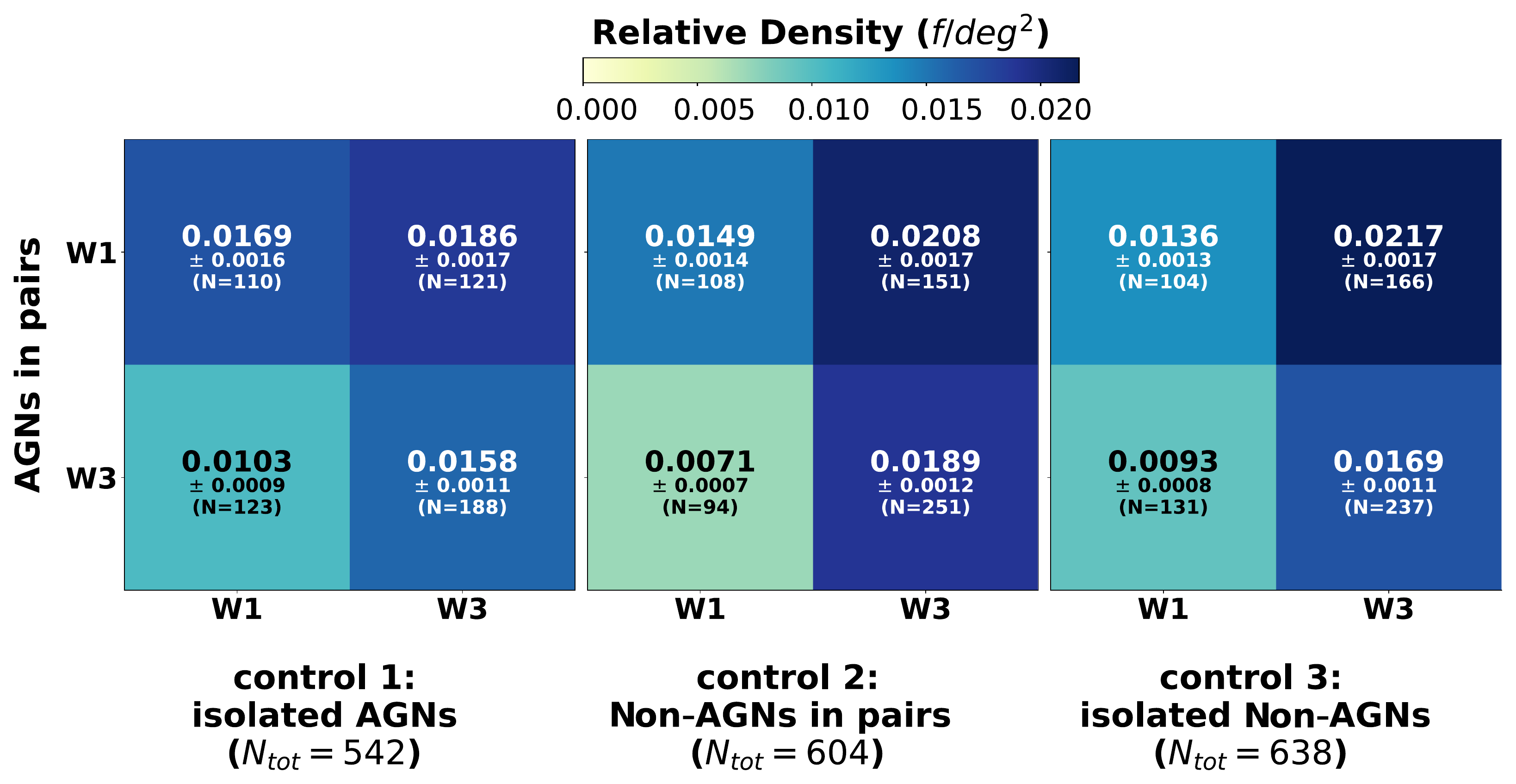}
    \caption{Confusion matrices showing the distribution of the three control samples across the W1 and W3 survey fields. Values indicate the relative density (fraction of the total sample per square degree) and the absolute number of objects (N) for each field intersection. Uncertainties represent 1$\sigma$ Poissonian errors.}
    \label{confusion_matrix}
\end{figure}

Control candidates were drawn from the entire master catalogue regardless of their specific field of origin. This cross-field-matching approach is not expected to introduce systematic biases, as the W1 and W3 fields exhibit high levels of consistency in terms of depth and number density. 
Control galaxies were selected through sampling without replacement, ensuring that each object only enters the control catalogues once and that the three control samples remain mutually disjointed, thereby preserving the statistical independence of the datasets. To quantitatively verify this, we analysed the distribution of the selected controls across the survey fields. Figure \ref{confusion_matrix} presents the confusion matrices for the three control samples, showing the relative density of objects (fraction of the total sample per square degree, f/$\rm deg^2$) for each field intersection. The densities are normalized by the total sample size and the respective field area to ensure comparability.
As illustrated in this figure, the spatial distribution of the controls is consistent across the three independent samples. The relative density values for the corresponding field intersections are statistically consistent within their $1\sigma$ Poissonian uncertainties. For instance, the primary field intersections exhibit a typical numerical consistency of $\sim$90\% across the three panels. This stability confirms that the selection function is not biased by field-specific variations in depth or ancillary data, ensuring that the control galaxies are comparably distributed relative to the reference AGNs-in-pairs sample.

Although up to three control galaxies were assigned to each AGN, the simultaneous matching in three parameters --- each with markedly different underlying distributions --- prevented us from obtaining exactly three controls for every object. The lower panels of Fig.~\ref{cont1} show the resulting distributions of redshift, absolute $i$-band magnitude, and local overdensity for the control samples. The number counts of the different samples are also presented. 
As the figure illustrates, the control samples successfully reproduce the distributions of the reference AGNs in pairs for all three parameters. We quantified the similarity between the control samples and the AGNs in pairs using Kolmogorov--Smirnov tests. In all cases, we obtained $p>0.05$, consistent with the null hypothesis that the control and reference samples are drawn from the same parent distributions.

Despite the successful matching in redshift, luminosity and global environment, the insets in Fig.~\ref{cont1} reveal systematic offsets in the stellar mass distributions ($M_{\star}$) between the reference and control samples. Both AGNs in pairs and isolated AGNs (\textsc{Control 1}) tend to exhibit higher stellar masses compared to the non-AGN populations (\textsc{Controls 2} and \textsc{3}). 
These discrepancies can be attributed to intrinsic physical differences in the host galaxies, but also to potential biases in the SED fitting procedure --- such as the age--dust--metallicity degeneracy --- where stellar mass estimates may be influenced by trends in colour and dust absorption. To mitigate these effects, a mass-re-weighting procedure was implemented in the following analysis (see Sect. \ref{sec:color_evolution}) to isolate the impact of environmental interactions and AGN activity from the fundamental influence of stellar mass on the observed galaxy properties.

\section{Results and discussion}
\label{results}
\subsection{Galaxy properties}
\label{sec:color_evolution}
In this section we explore different galaxy properties and their evolution with redshift. We consider AGNs in pairs and AGNs and non-AGNs  in the different control samples (\textsc{controls 1, 2, and 3}).  The associated errors in all figures of this section were calculated using bootstrap resampling techniques \citep{Barrow}.

 \begin{figure}[h!]
      \centering
      \includegraphics[width=0.48\textwidth]{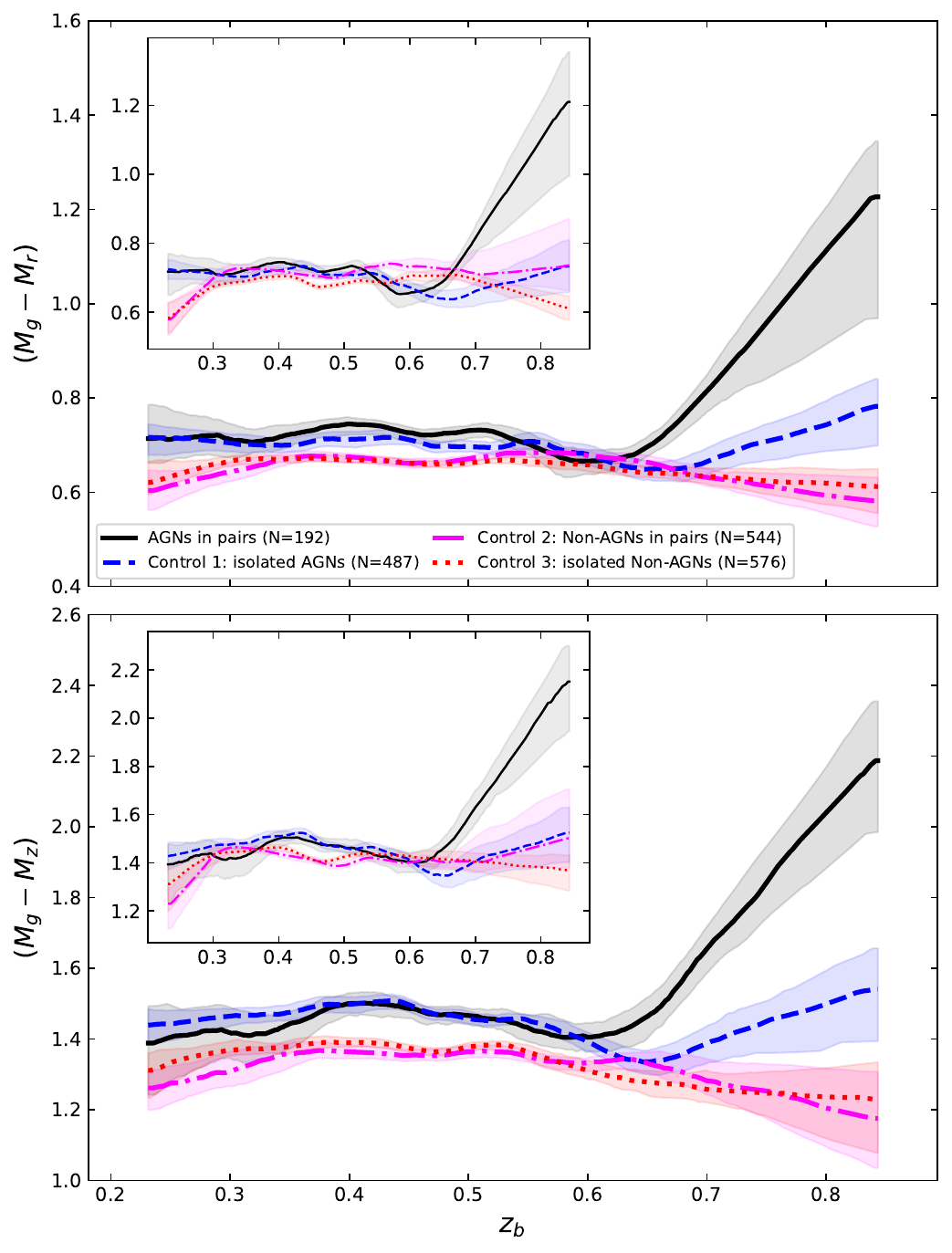}
  \caption{Colour index $(Mg-Mr)$ (\textit{top}) and $(Mg-Mz)$ (\textit{bottom}) as a function of redshift for four different samples of galaxies: AGNs in pairs (solid black line), control isolated AGNs (dashed blue line), control non-AGNs in pairs (dotted magenta line), and control isolated non-AGNs (dotted red line). The errors correspond to estimates from bootstrap resampling techniques.} 
  \label{fig:merged}
\end{figure}

\begin{figure*}[ht]
   \centering
   \includegraphics[width=.97\textwidth]{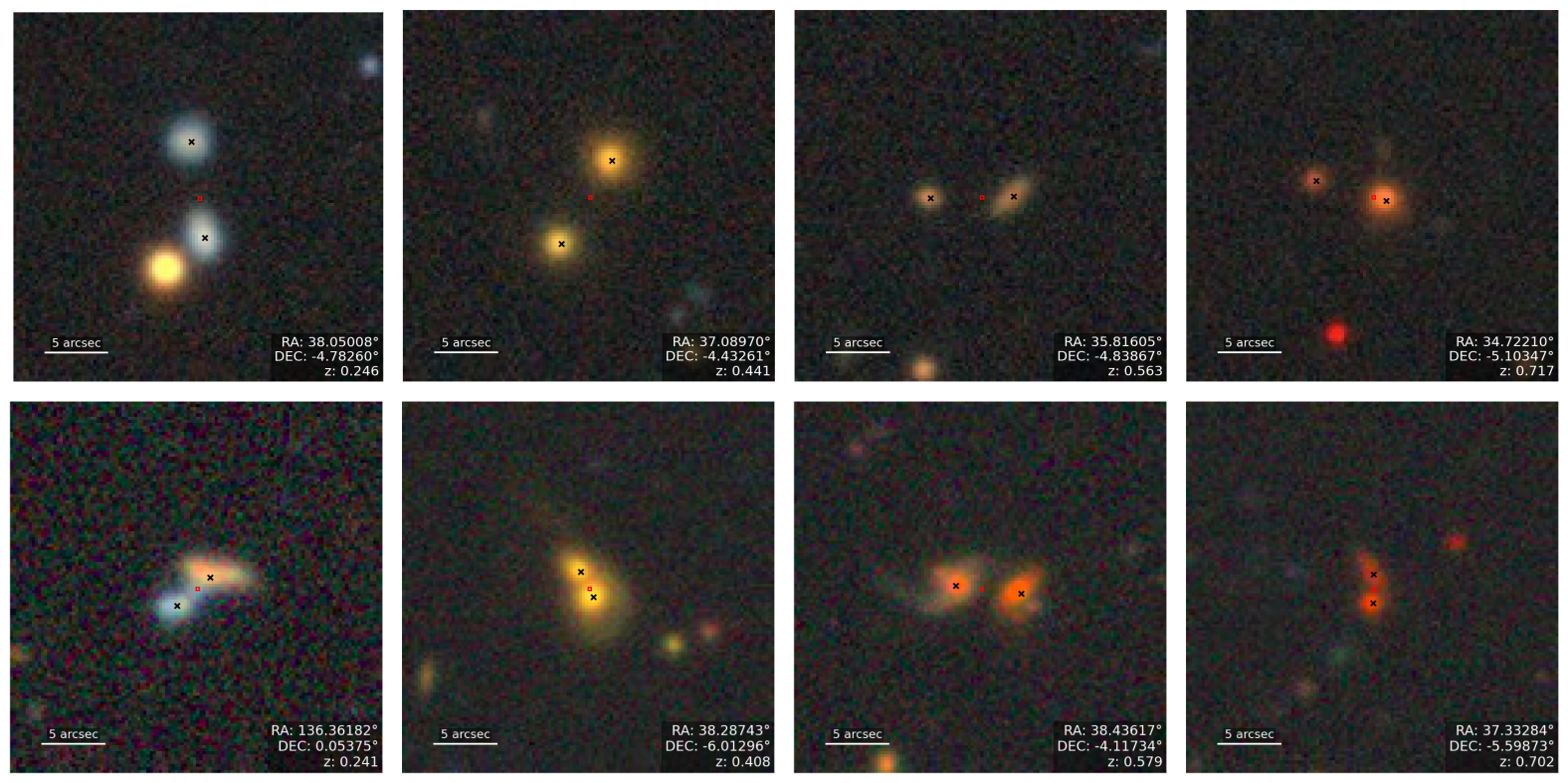}
   \caption{Examples of non-MD (\textit{top}) and MD (\textit{bottom}) pairs at different redshifts. Images are $30\arcsec \times 30\arcsec$ cut-outs from \url{https://www.legacysurvey.org/viewer/}, using data from DR10 (Credit: Legacy Surveys / D. Lang (Perimeter Institute)). The markers indicate the pair barycentre (open red circle), the positions of both pair members (black crosses), and a scale bar. We also overlaid the coordinates and redshift of the pair in the bottom right corner.}
   \label{rgbs}
\end{figure*}

Figure \ref{fig:merged} shows the evolution of the $(M_g - M_r)$ and $(M_g - M_z)$ colour indices as a function of redshift. To characterize the underlying trends in these galaxy properties, we employed a non-parametric smoothing approach based on locally weighted scatterplot smoothing (LOWESS). This method allows us to model the median behaviour of the population without assuming a specific functional form. The degree of smoothing was controlled by a fractional bandwidth parameter $f=0.3$, which determines the local bandwidth, $h$, as the distance to the $f \cdot N^{th}$ nearest neighbour. This adaptive bandwidth provides a stable estimation of the global trends while minimising sensitivity to local fluctuations and density variations across the redshift range. To avoid edge effects and ensure the stability of the central trend estimation, the analysis was restricted to a redshift range containing the central 90\% of the data, effectively clipping the extreme tails of the distribution where the low density of points could bias the local regression.

The insets in Fig. \ref{fig:merged} display the colour evolution after accounting for differences in the stellar mass distributions between the samples. This was achieved by a mass-re-weighting procedure, where weights were derived from the bin-by-bin ratio of the stellar mass distributions using a constant bin width of $\Delta \log M_{\star} = 0.125$ dex. This resolution was chosen to accurately resolve the observed offsets in the mass function peaks between the samples while maintaining sufficient statistics in each bin. 
For each control galaxy, $i$, falling within the $k$-th mass bin, the weight, $w_i$, was computed as

\begin{equation}
w_i \equiv w_k = \frac{N_{\text{ref},k} / N_{\text{ref, tot}}}{N_{\text{ctrl},k} / N_{\text{ctrl, tot}}},
\end{equation}

\noindent where $N_{\text{ref},k}$ and $N_{\text{ctrl},k}$ are the number of galaxies in the $k$-th stellar mass bin for the AGNs in pairs reference and the corresponding control subsamples, respectively. To ensure statistical stability, weights were clipped at the 95th percentile of the distribution. These weights were then incorporated into the LOWESS minimisation process.

Uncertainties for both the standard and weighted relations were estimated through bootstrap resampling with 100 realisations. The resulting distribution of smoothed curves was used to derive asymmetric confidence intervals defined by the $16^{th}$ and $84^{th}$ percentiles (equivalent to $1\sigma$ dispersion), represented as shaded regions around the median LOWESS relations.

Figure \ref{fig:merged} reveals several interesting evolutionary trends regarding the colour properties of AGNs in pairs and galaxies in the different control samples (\textsc{controls 1, 2, and 3}). In the standard relations (main panels), AGNs in pairs exhibit the reddest colours for photo-z $z_b > 0.6$, with an increasing trend towards higher redshifts. In contrast, non-AGN galaxies in the \textsc{control 2} and \textsc{3} samples show bluer colours across most of the redshift range and relatively flat trends. Isolated AGNs (Control 1) show an intermediate behaviour, remaining consistently bluer than their counterparts in pairs at high redshift. Notably, the differences between AGNs in pairs and all other control samples exceed the $3\sigma$ level for $z_b > 0.6$, suggesting that the pair environment exerts a specific influence on the host galaxy's colour properties.

In the mass-re-weighted insets, the characteristic redward trend for AGNs in pairs remains prominently visible. Even after mitigating the influence of the stellar mass distribution, the median relation for the reference sample maintains a clear separation from the control samples, with a significance level of $\geq 2\sigma$ in the highest redshift bins ($z_b > 0.6$). The persistence of this excess in colour after re-weighting indicates that the reddening in pair-hosted AGNs is not merely a consequence of higher stellar masses, but likely reflects physical processes associated with the pair environment and nuclear activity. These results suggest that the local pair environment and AGN activity contribute to systematic differences in galaxy colours, in addition to the fundamental dependence on stellar mass.

\subsection{Galaxy pair parameters}
\label{pair_prop}
In this section we explore different pair properties as a function of redshift. To this end, we examined the relation with redshift of different pair parameters such as the fraction of galaxies showing morphologically disturbed features, the projected separation between galaxies, $r_p$, and the radial-velocity difference, $\Delta V$. We considered the sample of pairs hosting AGNs and the control sample of non-AGNs in pairs (\textsc{control 2} defined in Sect. \ref{control_samples}).

In order to identify galaxy pairs with a higher probability of being physically associated and undergoing interactions, we complemented the pair selection with a morphological classification based on visual inspection. This subsample is expected to be less affected by projection effects, as the presence of morphological disturbances provides direct observational evidence of ongoing or recent interactions.

To identify morphologically disturbed galaxies in the pair sample, a visual inspection was conducted using, false-colour images from the DESI Legacy Survey DR10 \citep{Dey2019}. This approach was necessary as the PAUS survey, despite its strengths in providing high-quality NB photometry for galaxy classification, does not provide the suitable images required for this type of morphological analysis. Since the PAUS fields are fully contained within the Legacy Survey footprint, these publicly accessible images provided a consistent and ideal source for the visual characterisation of our galaxy pairs.

To facilitate accurate inspection, we generated customised $30\arcsec \times 30\arcsec$ RGB cutouts for each target pair. 
The images were retrieved from the Legacy Survey public cutout service\footnote{https://www.legacysurvey.org/viewer/} and processed using \texttt{Python} routines, ensuring uniform astrometric alignment and pixel scale. 
This approach enabled a systematic, reproducible, and visually consistent assessment of morphological disturbance signatures, such as tidal tails or bridges, in galaxies in our pair samples. Galaxies exhibiting these morphologically disturbed features were classified as MD galaxies, and pairs hosting at least one of these galaxies were defined as MD pairs.

Figure~\ref{rgbs} presents examples of MD pairs and non-MD pairs at different redshifts. 
Since the redshifts of the pairs shown in the top and bottom panels are comparable, the angular scale translates into a similar physical scale in each image, allowing a direct and meaningful comparison of the tidal features present in the systems.

\begin{figure}[h]
  \centering
       \includegraphics[width=0.45\textwidth]{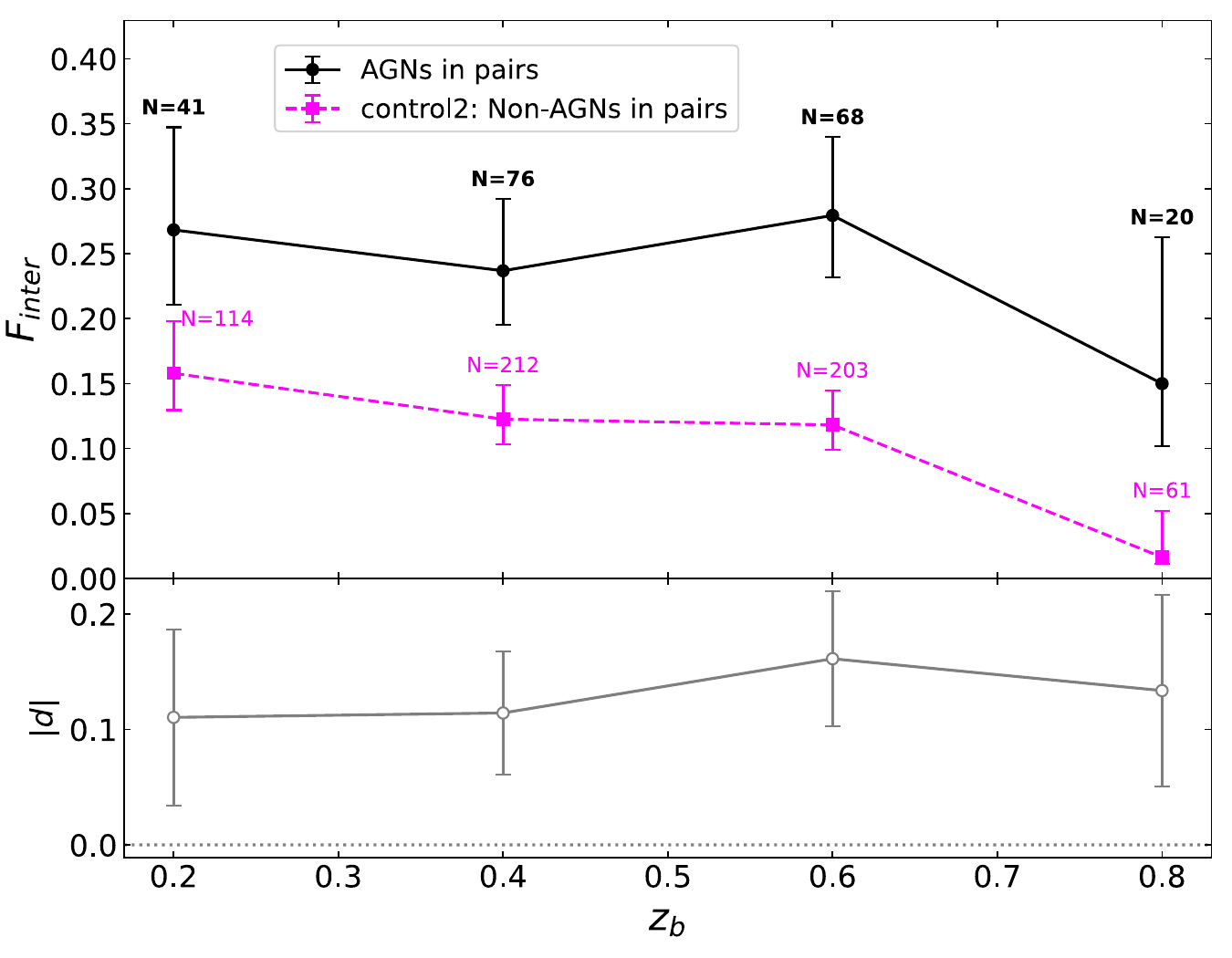}
    \caption{\textit{Top}: Fraction of galaxies in MD pairs ($F_{\mathrm{MD}}$) for the AGN pair sample (solid black line) and the \textsc{control~2} non-AGN pair sample (dot-dashed magenta line), shown in different redshift bins. The total number of galaxies in each sample per bin is listed in the figure. \textit{Bottom}: Absolute difference between the two fractions, $|d|$, as a function of redshift. Errors were estimated using the beta distribution quantile technique \citep{Cameron2011}.}

    \label{frac_inter}
\end{figure}

The visual classification was performed by the first author of this paper and refined iteratively until a 85\% agreement rate was achieved. The final results indicate that 24\% (51) of the AGNs in the pair sample exhibit morphological disturbance features, compared to 12\% (70 galaxies) in the corresponding \textsc{control~2} sample. 

Figure~\ref{frac_inter} displays the fraction of morphologically disturbed galaxies ($F_{\mathrm{MD}}$) for both samples. This fraction is defined as the number of MD galaxies identified in a given redshift bin relative to the total population of that bin for the AGN sample (solid black line) and the \textsc{control 2} sample (dot-dashed magenta line).

Both samples exhibit a general downward trend, with the fraction of MD galaxies declining towards higher redshifts. This trend could be the result of observational biases such as lower angular resolution and surface-brightness sensitivity, which hinder the detection of morphological disturbances at higher redshifts. However, since both samples are equally affected by these limitations, the relative comparison between them remains robust.
In the lower panel we present the absolute difference between the fractions. A consistent offset of $\sim$0.15 is observed between the two samples in all redshift bins.  
Therefore, AGNs in pairs are more likely to exhibit morphological disturbances by a fixed fraction ($|\rm d|\sim 0.15$) compared to the non-active galaxies in the \textsc{control 2} sample. This systematic excess of disturbed morphologies among the AGN population suggests that nuclear activity is preferentially triggered or sustained in systems showing clear observational evidence of ongoing interactions.
This consistent difference indicates that the link between morphological signatures of interaction and the presence of AGNs remains stable across the studied redshift interval.

\begin{figure}[ht]
   \centering
        \includegraphics[width=.48\textwidth]{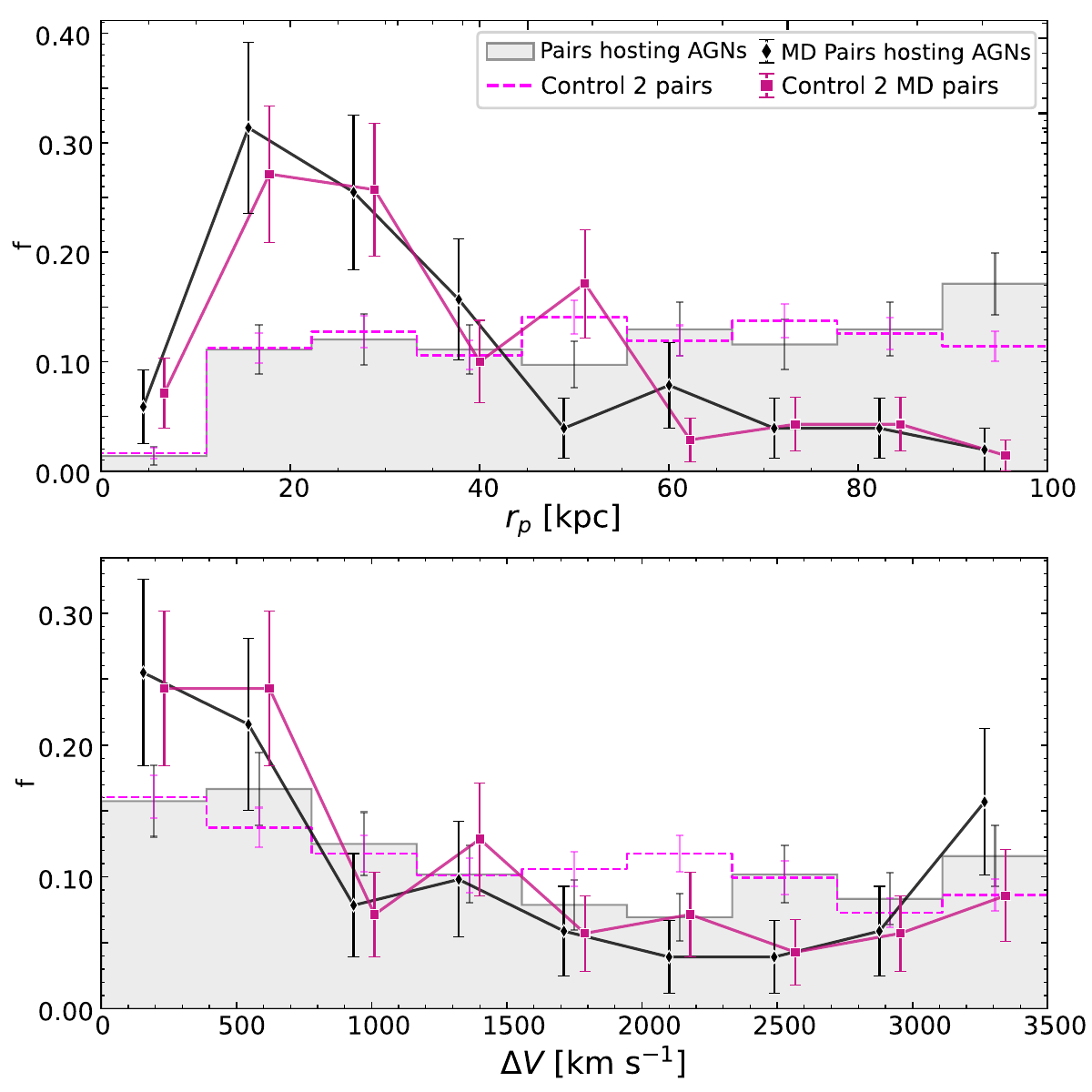}
     \caption{Normalised distributions of projected separation, $r_p$ (\textit{top}), and the radial-velocity difference, $\Delta V$ (\textit{bottom}), of pairs hosting AGNs (grey shaded histograms) and pairs hosting \textsc{control 2} non-AGNs (dashed magenta histograms). The distributions for MD pairs within each sample are shown with solid lines and markers. A small horizontal offset was applied to the MD markers to improve visibility. Error bars in all distributions represent Poissonian uncertainties.  }
    \label{rp_dV_dist}
\end{figure}

To study the pair configuration, we explored the trends of $r_p$ and $\Delta V$ for the sample of pairs hosting AGNs and pairs hosting \textsc{control 2} galaxies. Figure \ref{rp_dV_dist} shows the distributions of $r_p$ and $\Delta V$ of pairs in these samples, along with the corresponding MD subsamples. Both samples present similar distributions of the pair parameters. In the case of $r_p$, the drop at distances lower than 10$\kpc$ is associated with resolution limits in the catalogue.
For MD pairs, a general trend of having lower values of $r_p$ and $\Delta V$ can be appreciated for both pairs hosting AGNs and pairs hosting \textsc{control 2} non-AGNs.

For a joint characterisation of the pair parameters, we define the normalised radial distance, $R$, as 

\begin{equation}
R=\sqrt{ \left( {r_p \over r_{p\ \mathrm {max}}}\right)^2+\left({\Delta V\over dV_{\mathrm {max}}}\right)^2}
,\end{equation}
where $r_{p\ \mathrm {max}}=100\kpc$ and $\Delta V_{\mathrm  {max}}=3500\kms$. 

The definition of the normalised radial distance, R, serves to characterise the proximity of galaxies in pairs within the projected phase space. By combining $r_p$ and $\Delta V$ in a single dimensionless parameter, we can quantify the dynamical state of the pair more effectively than through independent cuts. Low values of R identify systems with both, small spatial separations and low relative velocities, which are the most likely candidates for physically bound systems experiencing significant mutual gravitational influence.

We considered four regions defined by the $R$ quantiles ($Q_1, Q_2, Q_3$), representing concentric shells of increasing dynamical distance in the normalised phase space:
\begin{itemize}
    
\item R1: $0<R\leq 0.54 $ 
\item R2: $0.54<R\leq 0.78$ 
\item R3: $0.78<R\leq 0.97 $ 
\item R4: $R>0.97$ 

\end{itemize}

\begin{figure}[ht]
   \centering
        \includegraphics[width=.35\textwidth]{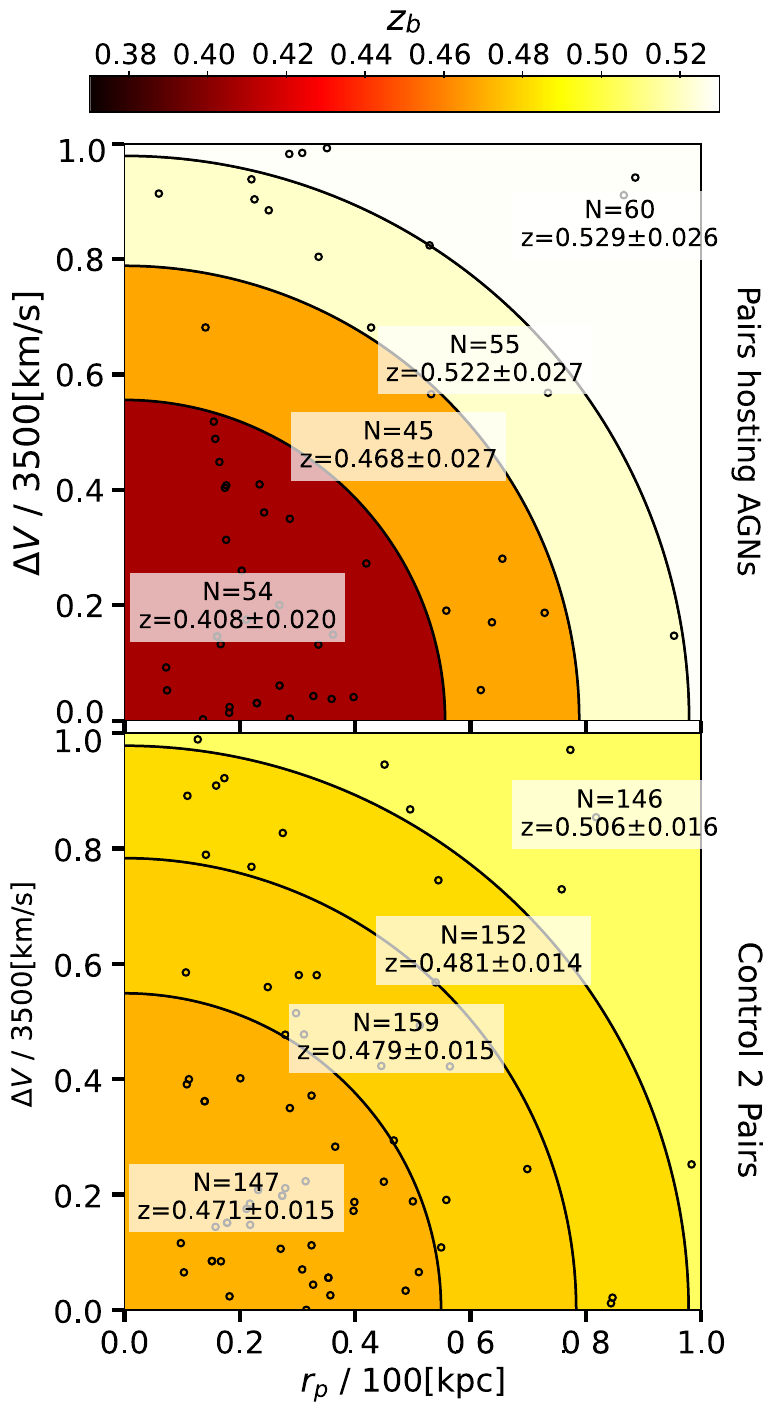}   
    \caption{Normalised radial-velocity difference ($\Delta V/\Delta V_{\mathrm{max}}$) versus normalised projected distance ($r_p/r_{p,\mathrm{max}}$) plane. The marked regions (R1--R4) correspond to the $R$ quantile segmentation and are colour-coded according to the average photo-z ($\langle z_b \rangle$) of the individual paired galaxies within each shell. The number $N$ indicates the total count of individual galaxies. Errors were calculated using bootstrap resampling. The upper panel shows the AGN pair sample, while the lower panel displays the corresponding \textsc{control 2} sample of non-AGN pairs.
    Dots represent the subsamples of MD pairs.}
    
    \label{rp_dV_plane}
\end{figure}

By partitioning the sample into $R$ shells, we can isolate the most compact and potentially bound systems from those at the outskirts of the interaction parameter space. Figure \ref{rp_dV_plane} shows the distribution of the samples in this normalised plane. Each region is colour-coded according to the mean photo-z ($\langle z_b \rangle$) of the individual paired galaxies within it. We find that for sample pairs hosting AGNs, the innermost shell ($R1$) is characterised by a lower mean redshift ($\langle z_b \rangle \approx 0.41$) compared to the periphery ($R4, \langle z_b \rangle \approx 0.53$), a gradient that is less pronounced in the \textsc{control 2} sample.

The subsamples of MD pairs are also plotted. We found a higher fraction of MD galaxies within R1 and R2 regions for both AGNs and control 2 pair samples. The percentages are 75\% compared to 50\% in the corresponding parent samples, linking these systems to close pairs and reinforcing the interpretation of pairs in these regions as physically bound systems.

The subsamples of galaxies belonging to MD pairs are also identified in this plane (see Fig.~\ref{rp_dV_plane}). We found a higher fraction of MD galaxies within R1 and R2 regions for both AGNs and control 2 pair samples. Specifically, approximately 75\% of all galaxies in MD pairs are located within these two innermost regions, whereas only 50\% of the galaxies in the corresponding parent samples reside in the same area. This preferential clustering at low $R$ values links MD pairs to close configurations, reinforcing the interpretation of systems in these regions as physically bound systems undergoing significant gravitational interaction.

To further explore the pair parameters, we define the \texttt{Proximity Index}, 
\begin{equation}
\texttt{PI}=\frac{{R}_{\mathrm {max}}-R} {{R}_{\mathrm {max}}},
\end{equation}

\noindent where $R$ is the normalised radial distance previously defined and ${R}_{max}$ is the value of $R$ when $r_p=r_{p\mathrm\ max}$ and $\Delta V=dV_{\mathrm  max}$
With this definition,  \texttt{PI} is directly related to the physical configuration of the pair, ranging from zero for loose pairs to one for close ones. 

\begin{figure}[ht]
   \centering
        \includegraphics[width=.48\textwidth]{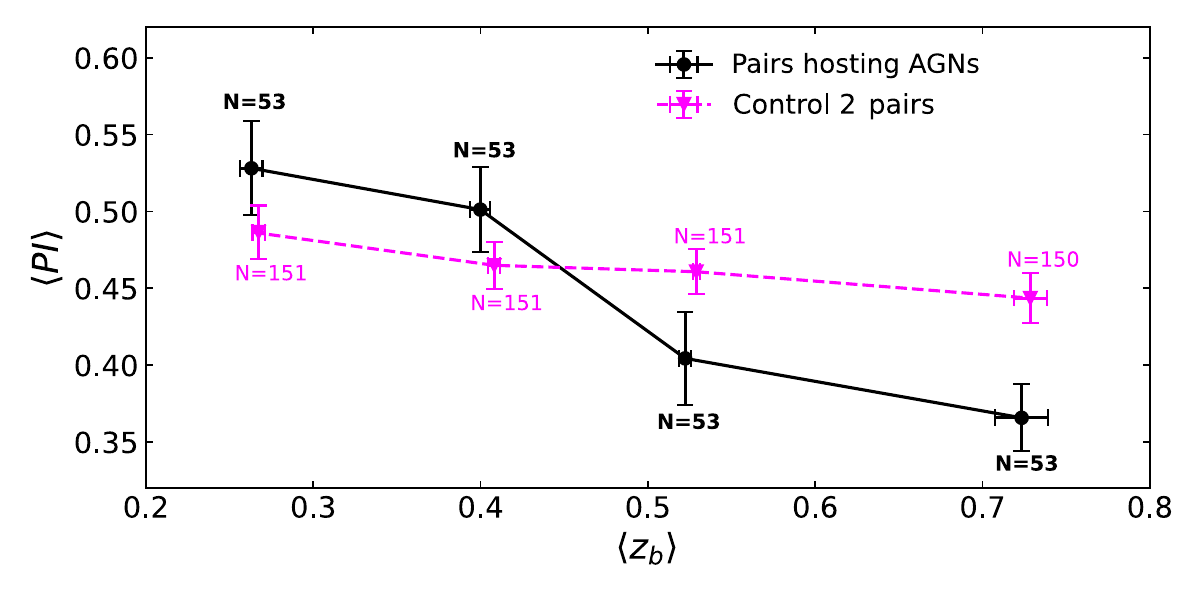}
    \caption{Average value of \texttt{PI} as a function of mean photo-z $\langle z_\mathrm b \rangle$. The solid black line represents pairs hosting AGNs, while the dot-dashed magenta line corresponds to pairs hosting non-AGNs in the \textsc{control 2} sample. The numbers of objects in each interval are listed. The errors were calculated using bootstrap resampling techniques.   }      
    \label{PI_zmed}
\end{figure}

In Fig.~\ref{PI_zmed} we plot the average \texttt{PI} ($\langle \mathrm{PI} \rangle$) as a function of the mean photo-z ($\langle z_\mathrm{b} \rangle$), calculated across four quantile bins in $z_\mathrm{b}$. This figure reveals a distinct behaviour in the proximity of systems hosting an AGNs compared to the \textsc{control 2} non-active pairs. Specifically, pairs hosting an active galaxy exhibit a clear dependency on redshift, where the average \texttt{PI} increases significantly (more that 3$\sigma$) towards lower redshifts. In contrast, the \textsc{control 2} sample displays a flatter trend, with the \texttt{PI} remaining relatively constant (within 1$\sigma$) across the same redshift range. This suggests a potential distinction in the dynamical state of pairs hosting AGNs compared to pairs hosting non-AGNs in the \textsc{control 2} sample, particularly as the system evolves towards more recent epochs.

\section{Summary and conclusions}
\label{conc}
We used high-precision photo-z from the PAUS survey in the W1 and W3 fields to carry out a detailed study of AGNs in galaxy pairs. The main goal of this work is to investigate how galaxy--galaxy interactions affect AGN triggering and how the physical properties of AGNs hosts in paired systems evolve with redshift.

To this end, we identified galaxy pairs following the methodology described in \citet{gonzalez2023pau}. Pairs were selected based on the following criteria: projected separation, $r_p < 100 \kpc$; radial-velocity difference, $\Delta V < 3500\kms $; and the requirement that at least one member has an absolute $r$-band magnitude brighter than $-17.5$. We further restricted the sample to galaxy pairs with a magnitude difference of $\Delta m < 2.5$ and applied an isolation criterion that excludes systems with a third galaxy within $300\kpc$ and $\Delta V < 3500\kms$ whose magnitude difference with respect to the central galaxy is also smaller than 2.5.
The resulting pair sample spans a stellar mass range of approximately $8 \lesssim \log(M_\star/M_\odot) \lesssim 11.5$, redshifts of $0.1 \lesssim z \lesssim 1.0$, and mass ratios consistent with the imposed magnitude difference threshold; thus, major and intermediate mergers were preferentially selected. Under these criteria, we identified a total of 11\,635 galaxy pairs in the PAUS survey.

Using PAUS galaxies, we identified AGNs by combining several multi-wavelength AGN catalogues, including selections based on MIR, optical, and X-ray data. This procedure yielded a catalogue of 7\,342 AGNs in W1+W3 PAUS fields spanning a wide redshift range, up to $z \sim 1$. By cross-matching this AGN catalogue with our galaxy pair sample, we identified 214 active galaxies residing in pair systems.

We also constructed three well-matched control samples comprising isolated AGNs, non-AGNs in pairs, and isolated non-AGNs. These samples were carefully selected to match the redshift, absolute magnitude, and local overdensity distributions of AGNs in pair systems. Such control samples are crucial for reliably identifying similarities and differences in galaxy properties when comparing populations selected through different methodologies. Using AGNs in pairs and different control samples, we analysed different galaxy properties and pair-specific parameters in order to investigate potential evolutionary trends as a function of redshift.

The main conclusions of this study are summarised below.

- The trends shown in Fig.~\ref{fig:merged} indicate that the colour indices $(M_g - M_r)$ and $(M_g - M_z)$ of AGNs in pairs evolve with redshift, becoming redder at higher redshifts, while isolated AGNs remain comparatively bluer. Control non-AGNs in pairs and AGNs in isolation show an almost flat trend of colour with redshift  ($M_g-M_r\sim 0.6$). The reddening of AGNs in pairs at higher redshifts could be linked to the peak of galaxy interactions and mergers at these epochs, potentially leading to an increased dust obscuration and star formation. Also, this might reflect changes in the fuelling mechanism of AGNs over cosmic time. 

- We found a statistically significant difference in $(M_g - M_r)$ and $(M_g - M_z)$ colours between AGNs in pairs and the control sample of isolated AGNs. This discrepancy is particularly prominent at higher redshifts ($z > 0.6$), where the separation exceeds $3\sigma$. When considering the mass-weighted distributions, the difference remains at the $2\sigma$ level. This indicates that the observed colour offsets are not primarily driven by systematic differences in stellar mass, although possible deviations at the highest masses cannot be ruled out. These results suggest that interaction-related processes in pairs may influence the photometric properties of AGN hosts. The redder colours observed in AGNs in pairs could indicate increased dust extinction, partial star-formation quenching, or a distinct AGN population associated with interacting systems.

- After a visual inspection performed using the DESI Legacy Survey images, we found that the fraction of morphologically disturbed (MD) galaxies is consistently higher in pairs hosting AGNs (24\%) compared to the \textsc{control 2} non-AGN pairs (12\%; see Fig.~\ref{frac_inter}). Although this fraction decreases with increasing redshift in both samples --- likely due to observational limitations such as reduced angular resolution and surface brightness sensitivity at earlier epochs --- the relative excess remains significant across the studied range. This trend supports a scenario in which galaxy interactions, as traced by visible morphological disturbances, play a key role in the triggering of nuclear activity.

- The distribution of the AGNs in pairs and \textsc{control 2} non-AGNs in the $r_p$--$\Delta V$ plane (Fig.~\ref{rp_dV_plane}) reveals that MD pairs are preferentially located at small projected separations and low relative velocities. This concentration in the innermost regions of the parameter space ($R1$ and $R2$) supports their classification as physically bound systems undergoing strong gravitational interactions. This trend is particularly pronounced for AGNs in pairs, which exhibit a higher density of MD galaxies in these compact configurations compared to the non-active \textsc{control 2} sample.

- Furthermore, \texttt{PI}, which quantifies the dynamical state of the systems, reveals a clear evolutionary trend for pairs hosting AGNs, transitioning from looser to tighter configurations towards lower redshifts. Specifically, the difference in the average \texttt{PI} between the highest and lowest redshift bins for these systems exceeds $3\sigma$. In contrast, pairs hosting \textsc{control 2} non-AGNs exhibit no significant evolution, with the difference between the same redshift extremes remaining within $1\sigma$. These results suggest a redshift-dependent distinction in the dynamical configuration of AGN-hosting pairs, possibly indicating that the efficiency of interactions in triggering nuclear activity increases as systems become more compact at later epochs (see Fig.~\ref{PI_zmed}).

The results of this work leverage PAUS unique photo-z precision in providing evidence for the link between galaxy--galaxy interactions and AGNs activity, highlighting their evolving influence across cosmic time. Although our findings shed light on the physical characteristics of AGNs in pairs and their evolution with redshift, several exciting topics should be further explored.
In future research we aim to focus on the relationship between AGNs and galaxy--galaxy interactions considering different properties of galaxies, such as morphology, star formation rates, and nuclear feeding mechanisms. Extending our analysis to deeper fields and complementing photometric data with spectroscopic monitoring will be critical to improving our understanding of the connection between galactic evolution and AGN activity. In this context, the synergy with upcoming large surveys, such as DESI \citep{DESI2025} and LSST \citep{LSST2019}, will offer different opportunities to complement the present work. Furthermore, exploring whether hydrodynamic simulations that incorporate AGN feedback, such as IllustrisTNG \citep{Nelson2019} or EAGLE \citep{Schaye2015}, can successfully reproduce the results of this work may help us to gain a deeper understanding of the underlying physical processes governing the co-evolution of galaxy pairs and AGNs in the Universe.
\section*{Data availability} 

The PAUS public data, including the raw and the reduced images and the AGN catalogue used in this work can be accessed via CosmoHub \citep{Carretero2017,Tallada2020}, through this link \url{https://pausurvey.org/public-data-release/}.
The samples used in this work are only available in electronic form at the CDS via anonymous ftp to cdsarc.u-strasbg.fr (130.79.128.5) or via http://cdsweb.u-strasbg.fr/cgi-bin/qcat?J/A+A/.

\begin{acknowledgements}
FD and SA acknowledge the support of the Consejo Nacional de Investigaciones Cient\'ificas y T\'ecnicas de la Rep\'ublica Argentina (CONICET) and Secretar\'ia de Ciencia y T\'ecnica de la Universidad Nacional de San Juan. 
M.S. acknowledges support by the State Research Agency of the Spanish Ministry of Science and Innovation under the grants 'Galaxy Evolution with Artificial Intelligence' (PGC2018-100852-A-I00) and 'BASALT' (PID2021-126838NB-I00) and the Polish National Agency for Academic Exchange (Bekker grant BPN/BEK/2021/1/00298/DEC/1). This work was partially supported by the European Union's Horizon 2020 Research and Innovation program under the Maria Sklodowska-Curie grant agreement (No. 754510). 
DNG acknowledges support from the European Research Council (ERC) under the European Union’s Horizon 2020 research and innovation program with Grant agreement No. 101053992.
FR acknowledges support from the ICTP through the Junior Associates Programme 2023-2028. 
FR and AO would like to thank the Secretaría de Ciencia y Tecnología of the Universidad Nacional de Córdoba for their support. 
IVD-P acknowledges funding by NASA under the CRESST II program.
ME acknowledges support from the Spanish Plan Nacional project PID2023-152069NA-I00 funded by MCIN/AEI/10.13039/501100011033
JC acknowledges support from the grant PID2021-123012NA-C44 funded by MCIN/AEI/ 10.13039/501100011033 and by "ERDF A way of making Europe".
FJC acknowledges support from the Spanish Plan Nacional project PID2022-141079NB-C31.
JG-B acknowledges support from the Research Project PID2024-159420NB-C43 [MICINN-FEDER], and the Centro de Excelencia Severo Ochoa Program CEX2020-001007-S at Instituto de F\'isica Te\'orica.
HH is supported by a DFG Heisenberg grant (Hi 1495/5-1), the DFG Collaborative Research Center SFB1491, an ERC Consolidator Grant (No. 770935), and the DLR project 50QE2305.
MM acknowledges support from the Spanish Ministry of Science and Innovation through the project PID2021-124243NB-C22. This work was partially supported by the program Unidad de Excelencia Mar\'ia de Maeztu CEX2020-001058-M.
CP acknowledges support from the Spanish Plan Nacional project PID2022-141079NB-C32.
PR acknowledges the support by the Tsinghua Shui Mu Scholarship, the funding of the National Key R\&D Program of China (grant no. 2023YFA1605600), the science research grants from the China Manned Space Project with No. CMS-CSST2021-A05 and the Tsinghua University Initiative Scientific Research Program (No. 20223080023). 
ES acknowledges support from the Spanish Plan Nacional projectPID2021-123012NB-C42 funded by MCIN/AEI/10.13039/501100011033.
This work has made use of CosmoHub, developed by PIC (maintained by IFAE and CIEMAT) in collaboration with ICE-CSIC. It received funding from the Spanish government (grant EQC2021-007479-P funded by MCIN/AEI/10.13039/501100011033), the EU NextGeneration/PRTR (PRTR-C17.I1), and the Generalitat de Catalunya. 
This research made use of \texttt{Astropy}, (http://www.astropy.org) a community-developed core \texttt{Python} package for Astronomy \citep{astropy:2013, astropy:2018} and \texttt{TOPCAT}- Tool for OPerations on Catalogues And Tables \citep{Topcat}.

\end{acknowledgements}

\bibliographystyle{aa.bst}
\bibliography{bib}{}

\begin{appendix}

\section{Testing zphot quality for AGNs}
\label{sec:photo-z_accuracy}
The aim of this section is to compare the performance of weighted version of the {\sc BCNZ2} (hereafter {\sc BCNZ2w}) and {\sc DEEPz} for non-AGNs and AGNs.
As described in \citet{Eriksen_bcnz}, for {\sc BCNZ2w} the observed flux of a galaxy is fitted from a linear combination of redshift-dependent templates, which take into account both the continuum and the emission line terms of the \gls{sed}, for both NB and BB data. The probability redshift distribution of an object is computed as follows:

\begin{equation}\label{eq:pz}
    p(z) \propto \int _{\alpha_{1}\geq 0} \textrm{d}\alpha_{1} \cdot\cdot\cdot \int _{\alpha_{n}\geq 0} \textrm{d}\alpha_{n} e^{-0.5\chi^{2}(z, \alpha)} p_{\mathrm{prior}}(z, \alpha),
\end{equation}

\noindent where $\alpha_{i}$ corresponds to the amplitude of template $i=1,\dots,n$, $p_{\mathrm{prior}}$, to the priors, and $\chi^{2}$ is

\begin{equation}\label{eq:chi2}
    \chi^{2}(z, \alpha) = \sum_{i, \text{NB}}\left ( \frac{\phi_{i}^{\text{obs}} - l_{i}k\phi_{i}^{\text{model}}}{\phi_{\textrm{err},i}} \right )^{2} + \sum_{i, \text{BB}}\left ( \frac{\phi_{i}^{\text{obs}} - l_{i}\phi_{i}^{\text{model}}}{\phi_{\textrm{err},i}} \right )^{2},
\end{equation}

\noindent with $\phi_{i}^{\text{obs}}$ and $\phi_{\textrm{err}, i}$ the observed flux and its error in band $i$, respectively, while $\phi^{\text{model}}$ corresponds to the modelled flux. Finally, $k$ is a calibration parameter that accounts for the different amplitudes per galaxy in the NB and BB fluxes, while $l$ is another calibration parameter per band that accounts for the offset between the observed flux and the best-fitting model. The best-fitting template is obtained via the minimization of the $\chi^{2}$ (Eq.~\ref{eq:chi2}), within a redshift range of [0, 2], so that a redshift probability distribution $p(z)$ (Eq.~\ref{eq:pz}) can be assigned for each object.

On the other hand, {\sc DEEPz} was described and tested in the COSMOS field by \citet{Eriksen_2020}. This method consists of an autoencoder with ten layers and 250 nodes in both the encoder and decoder. Each layer applies linear transformations followed by ReLU non-linearities, batch normalization, and a 2\% dropout rate, except for the last three layers.
The autoencoder processes galaxy flux ratios, and its latent variables are combined with the galaxy flux ratios as input for a mixture density network (MDN), which represents the redshift distribution as a linear combination of 10 normal distributions. A 1\% dropout rate is applied after all linear layers in this network. The photo-z is obtained as the mode of the distribution. 

\begin{figure}[h]
\centering
        \includegraphics[width=.45\textwidth]{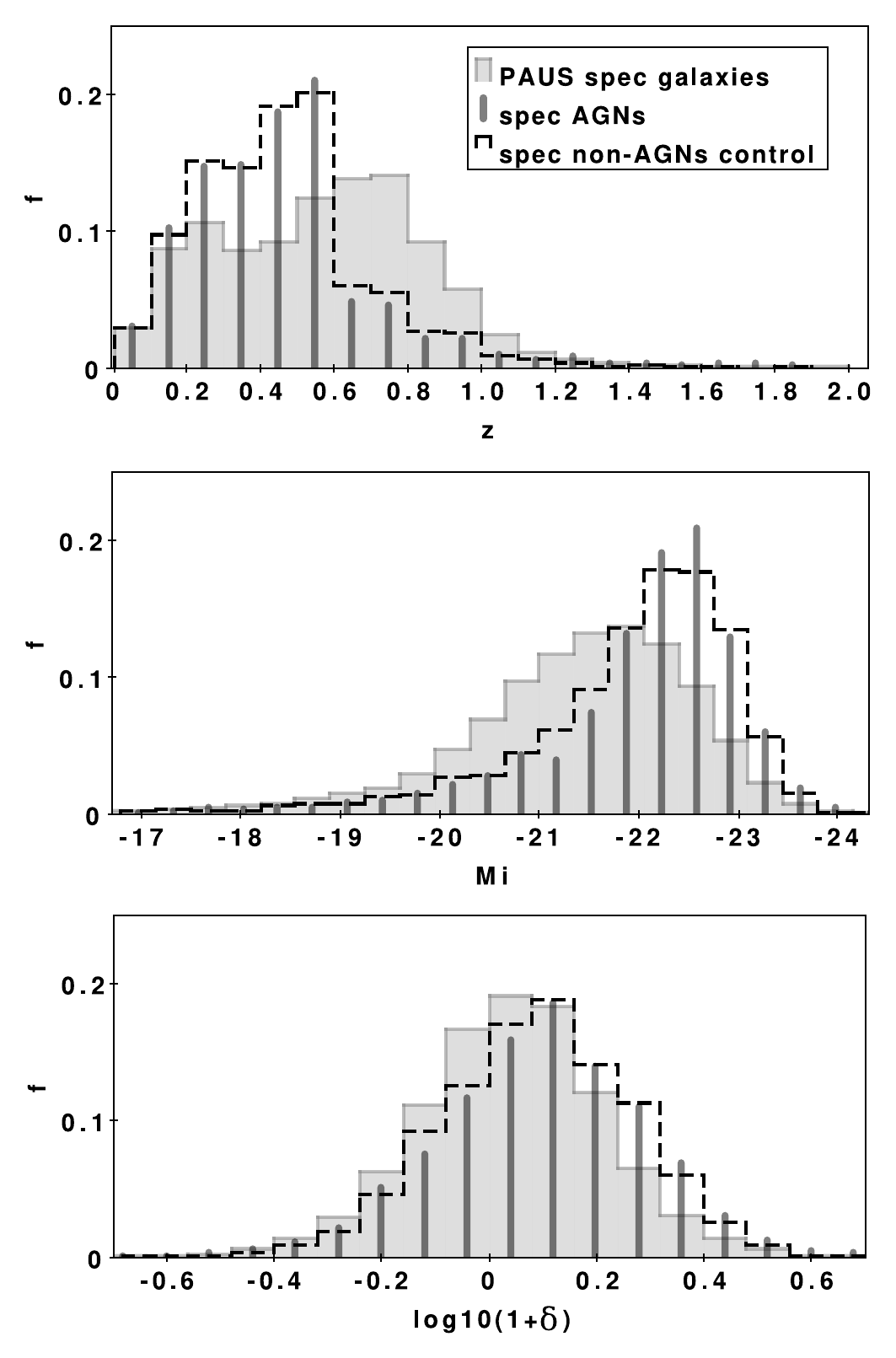}
    \caption{Normalised distributions of redshift (zs; \textit{top}), $i$-band absolute magnitude (Mi; \textit{middle}) and overdensity ($\delta$; \textit{bottom}) of spectroscopic PAUS galaxies, AGNs and the control sample.
    }
    \label{spec_control}
\end{figure}

DEEPz considers a sample of active galaxies in the training set used to compute photo-z, but BCNz code has no AGN templates. Therefore, it is crucial for this work to check the performance of the photometric redshift estimates for AGNs. To this end, we consider spectroscopic galaxies in PAUS data finding 58\,466 regular galaxies with spectroscopic redshift and 3267 spectroscopic AGNs. Figure \ref{spec_control} shows the distributions in redshift ($z_\mathrm s$), $i$-band absolute magnitude ($M_i$) and overdensity ($\delta$) for spectroscopic PAUS galaxies and AGNs. It can be seen that there are differences in the distribution of these parameters for regular galaxies and AGNs. 
Therefore, it is crucial to build suitable control samples before testing the photo-z accuracy. To this end, we match spectroscopic non-AGNs  in the PAUS sample to AGNs. We consider as control galaxies the nearest neighbours in the ($z_\mathrm s$, $M_i$, $\delta$) space. For some AGNs it was not possible to select a control galaxy, but we built a control sample of 3741 galaxies for 3758 AGNs (99.5\%). The results are shown in Fig. \ref{spec_control}, in all cases we found a Kolmogorov-Smirnov test p-value greater than 0.05, for the null hypothesis that the samples were drawn from the same distribution.

Using these controlled samples, we tested the photo-z performance by calculating the $\sigma_{\mathrm 68}$, the outlier rate and the bias for AGNs and control galaxies. 

Considering $z_\mathrm b$ and $z_\mathrm s$, the photometric and spectroscopic redshifts, respectively, and 
\begin{equation}
\Delta z=\frac{{z}_\mathrm b-z_\mathrm s}{{1+z_\mathrm s}}
\end{equation}

\noindent we define $\sigma_{68}$ as
\begin{equation}
\sigma_{68}=\mathrm P[84]-\mathrm P[16]
\end{equation}

\noindent where $\mathrm P[x]$ corresponds to the percentile x of the $\Delta z$ distribution. 
For the outlier fraction, we consider as outliers those objects with

\begin{equation}
|\Delta z|> 0.1
.\end{equation}

The bias is defined as the median of the difference
\begin{equation}
\mathrm {bias}=\mathrm {med}(z_\mathrm b-z_\mathrm s)
.\end{equation}

\noindent This quantity shows the systematic difference between the spectroscopic and photo-z.

\begin{figure*}
\centering
    \includegraphics[width=.9\textwidth]{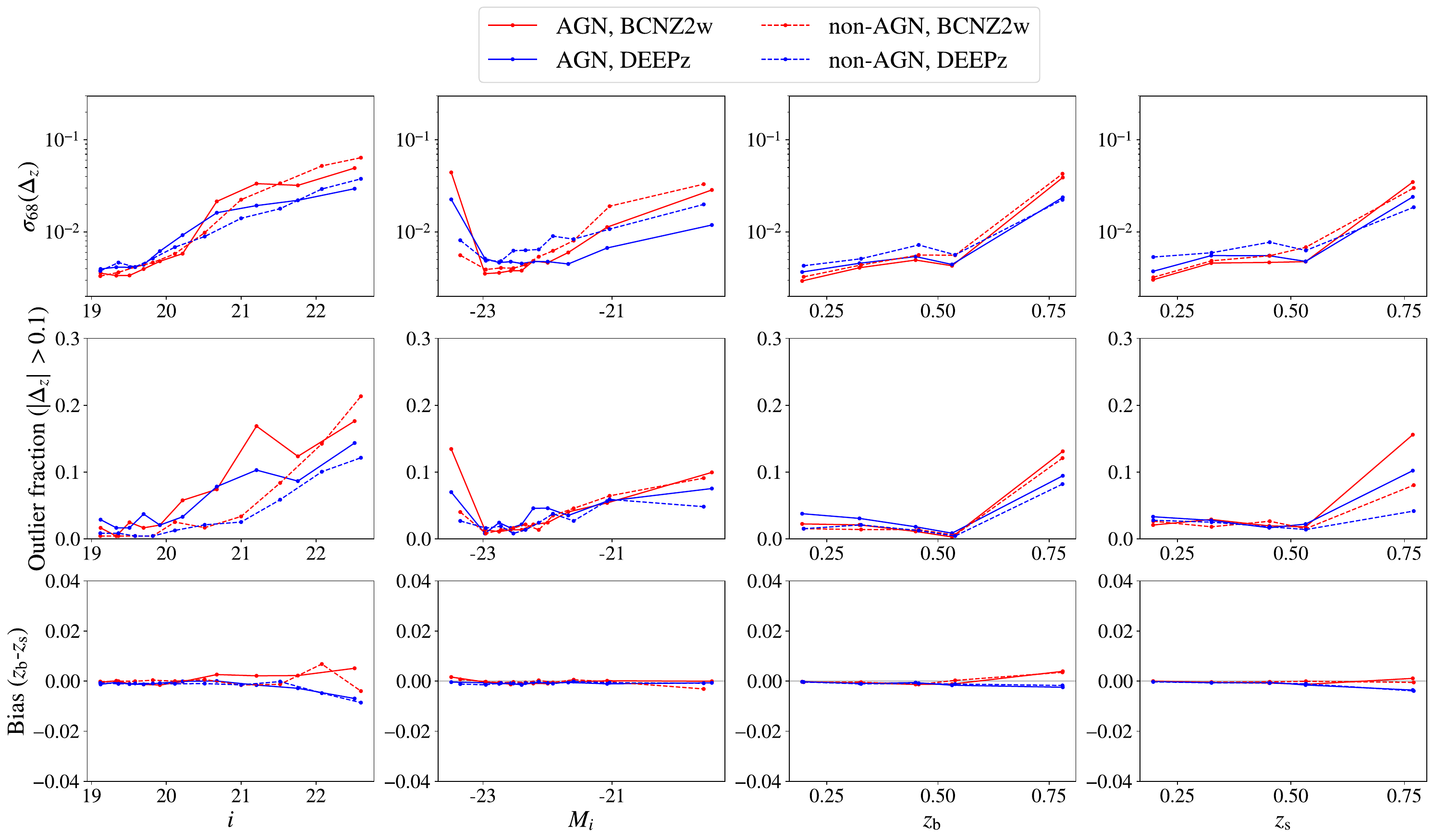}
    \caption{Photo-z accuracy for BCNZ2w and DEEPz for the AGNs and the non-AGN control sample. We show the $\sigma_{68}$ (\textit{top panels}), the outlier fraction (\textit{middle}) and the bias (\textit{bottom}). We consider the performance of these metrics as a function $i_{AB}$-band apparent ($i$) and absolute ($M_{i})$ magnitudes, as well as the trends for the photometric ($z_\mathrm b$) and the spectroscopic redshifts ($z_\mathrm s$).}
    \label{photo_z_metrics}
\end{figure*}

We use these metrics to check the performance of both BCNZ2w and DEEPz, photo-z estimates. The results are shown in Fig. \ref{photo_z_metrics} as a function of $i_{ab}$ magnitude and redshift. A decrease in precision and an increase in the outliers fraction can be observed towards fainter galaxies and higher redshifts.
We found similar trends in accuracy for both estimates in the overall sample; however, a difference in precision is observed for magnitudes fainter than $i\sim 20$ and $M_i\sim -21$ and for $z>0.6$. In particular, for galaxies with $i>20$, DEEPz demonstrates a reduction of 40\% and 31\% in $\sigma_{68}$ and outlier fraction, respectively, compared to BCNZ2w. These results agree with those of \cite{Daza-Perilla2025}, where the photo-z obtained using BCNz2 and DEEPz were compared for galaxies in the W1, W3, and G09 fields.

\begin{figure}[h]
\centering
    \includegraphics[width=.48\textwidth]{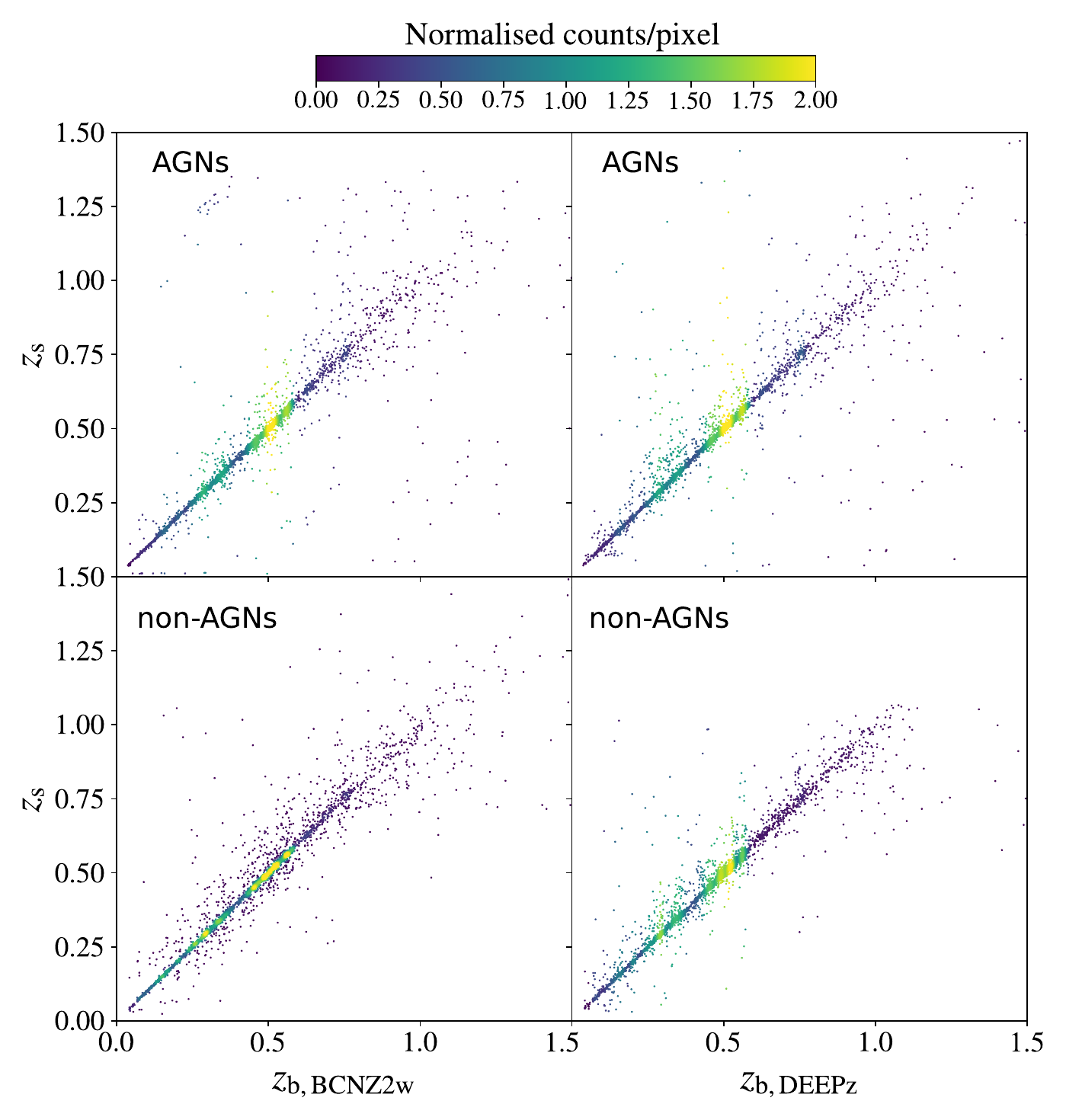}
    \caption{$z_\mathrm b$ versus $z_\mathrm s$ for the AGNs (\textit{left}) and the non-AGN control sample (\textit{right}). In the top panels we show the estimates of BCNZ2w and in the bottom those of DEEPz. The colour bar represents the normalised (over the median) counts per pixel.}
    \label{zb_vs_zs}
\end{figure}

\begin{figure}
\centering
    \includegraphics[width=.40\textwidth]{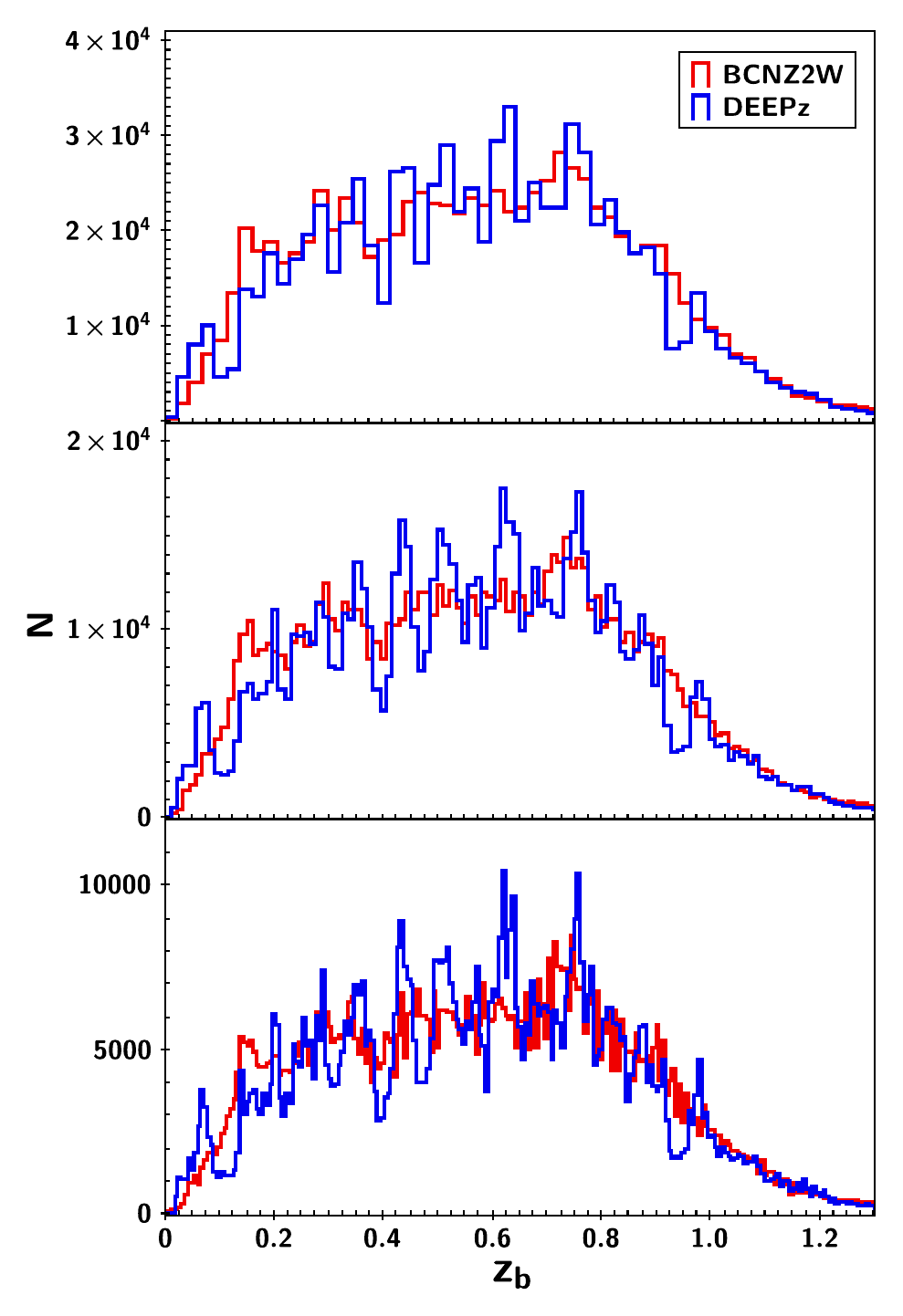}
    \caption{Redshift distribution for BCNZ2w and DEEPz photo-z estimates of PAUS galaxies. From top to bottom the panels show the photo-z distribution considering a bin that corresponds to 2, 1 and 0.5 times the maximum velocity difference between pair galaxies $\Delta V=3500\ \mathrm {kms^{-1}}$.}
    \label{zdists}
\end{figure}

In Fig. \ref{zb_vs_zs}, we present the spectroscopic versus photometric redshift relation for AGNs (top) and control galaxies (bottom) for BCNZ2w (left) and DEEPz (right), coloured by the normalised (over the median) counts per pixel. This plot depicts similar behaviours for both estimates, reinforcing the lack of significant differences between the performance of photo-z for AGNs and regular galaxies in the control sample. 
Therefore, unbiased results are expected in the implementation of photo-z in our study and both, BCNZ2w and DEEPz photo-z are suitable for our study of AGNs. However, the estimates of {\sc DEEPz}  photo-z present a discrete behaviour (see Fig. \ref{zdists}) that can introduce important noise in the pair identification for pairs closer than the discrete peak values. This noise is introduced in DEEPz training given that the observational sample used for training contains a higher density of examples near the peaks seen in the distribution, which biases the model toward those discrete values. 

Furthermore, the identification of close pairs using DEEPz and BCNz2 was evaluated in \cite{Daza-Perilla2025} finding that only 40\% of the pairs identified are consistent between the two methods. 
For these reasons, in the present work, we decided to use BCNZ2w. Additionally, these estimates were previously considered in the construction of pair catalogues with PAUS data \citep{rodriguez2020close, gonzalez2023pau}, making these estimates suitable for the study of AGNs in galaxy pairs aimed at this work.

\section{Testing the pair catalogue}
\label{Pairstest}
To test the quality of the pair catalogue, we decided to study a sample of spec-spec pairs. From the pair sample comprising 16\,824 systems (see Sect. \ref{paircat}), we selected 385 pairs with both galaxies having spectroscopic available redshifts from \citet{navarro_photoz} in our PAUS master catalogue; we call this sample `spec-pairs'.

\begin{figure}[h]
   \centering
        \includegraphics[scale=.5]{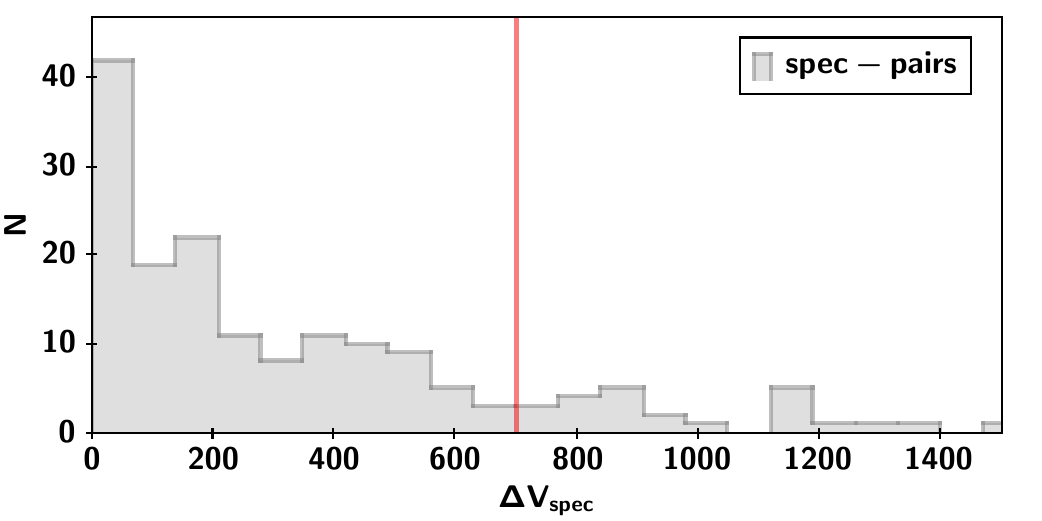}
    \caption{ Distribution of the spectroscopic radial velocity difference between galaxies in the spec-pairs. The vertical line represents the value used to define physical pairs. The distribution has a large tail up to 50000 $\mathrm {kms^{-1}}$ but was truncated to 1500 $\mathrm {kms^{-1}}$ to improve visualization.
    }
    \label{dVss}
  \end{figure}

The radial velocity difference between galaxies in the spec-pairs sample is shown in Fig.~\ref{dVss}. The $\Delta V_{\mathrm {spec}}$ values reach $50\,000 \kms$, evidencing the presence of spurious pairs that were removed prior to the construction of the control samples, ensuring that projection effects do not bias the results presented in this work.

We define a sample of physical pairs considering $r_p < 100 \kpc$ and $\Delta V_{\mathrm {spec}} < 700\kms$. This limit is higher than the usual value $\Delta V < 500$~$\mathrm {km\,s^{-1}}$, widely used to select galaxy pairs \citep[e.g.][]{Ellison2008,Patton2011,Tempel2018b,Duplancic2018,Alonso2024}. Nevertheless, the spectroscopic redshifts used in this work are derived from \citet{navarro_photoz}, who used different redshift catalogues with precision ranging from 50~$\mathrm {km\,s^{-1}}$ to 150~$\mathrm {km\,s^{-1}}$. Moreover, from Fig.~\ref{dVss}, we can see that the distribution of radial velocity difference drops around this value.

\begin{figure}[h]
   \centering
        \includegraphics[scale=.45]{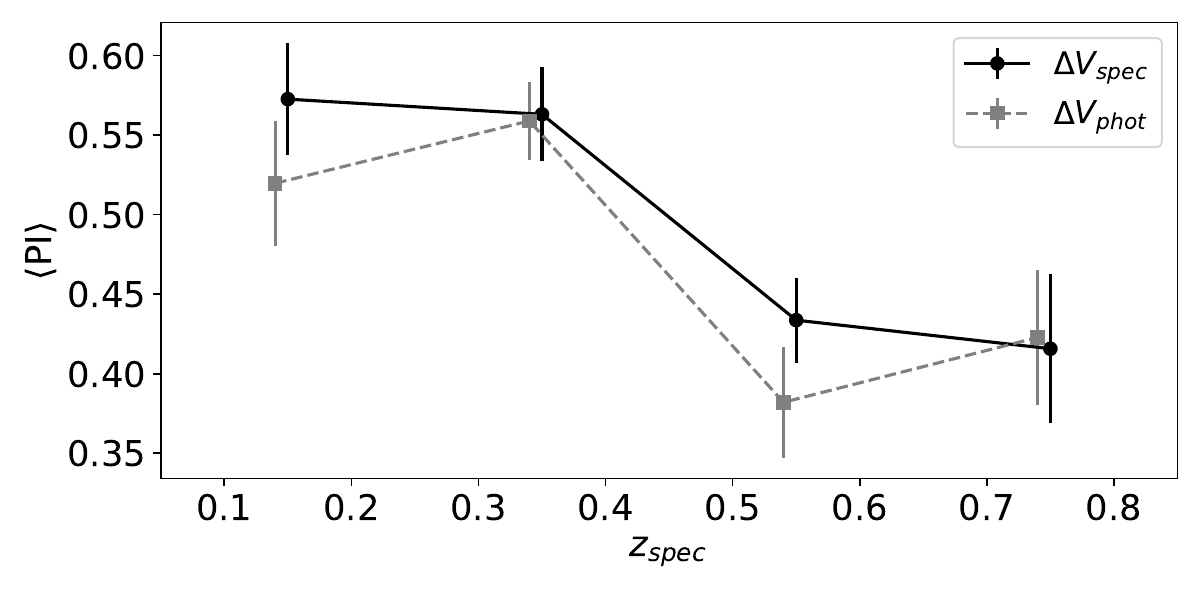}
    \caption{Average \texttt{PI} $\langle \texttt{PI} \rangle$ as a function of pair spectroscopic redshift for the sample of physical pairs. Solid lines correspond to the index estimation using spectroscopic radial velocity differences ($\Delta V_{\mathrm{spec}}$) and dashed lines to \texttt{PI} calculated with photo-z radial velocities ($\Delta V_{\mathrm {phot}}$). 
    }
    \label{PI_spec}
  \end{figure}

The sample of physical pairs represents 60\% of the spec-sample. Under this result, we conclude that  contamination of our pair catalogue can reach values of up to 40\%. The spec-spec pair analysis also provides an estimate of the level of contamination in our photometrically selected pair catalogue. This contamination level is higher than the $\sim$20\% reported by \citet{rodriguez2020close} using PAUS-based mock catalogues, where the pair selection was optimised to maximise purity and completeness. We note, however, that this estimate should be regarded as an upper limit, since the spec-spec subsample is not statistically representative of the full photometric pair catalogue.

To further study the quality of the photometric pairs, we studied the \texttt{Proximity Index} (\texttt{PI}) defined in Sect.~\ref{pair_prop}. We considered photometric and spectroscopic radial velocity differences to calculate this index for physical pairs. The results are shown in Fig.~\ref{PI_spec}, where we plotted average \texttt{PI} values as a function of the pair spectroscopic redshift. From this figure, we can appreciate that the \texttt{PI} values calculated with photometric and spectroscopic redshifts are consistent within the error bars. Therefore, we conclude that photo-z are a reliable tool to characterise the main parameters of our pairs.

\section{Comparison of different AGN selection methods}
\label{testing_AGNs}

To study the overlap and mutual consistency of the AGN classification methods used in this work, we used the UpSet plot technique \citep{Lex2014upset} to analyse the intersections among multiple sets. Our dataset includes four AGN classifications: MQS-AGNs, BPT-AGNs, MIR-AGNs, and X-ray AGNs. For each possible combination of these classifications (e.g. sources identified simultaneously as MQS- and MIR-AGNs, or as MQS-, BPT-, and X-ray AGNs), we computed the intersection size by counting galaxies that satisfy all criteria in each specific combination. This approach allows a quantitative comparison of the degree of overlap and complementarity among the different AGNs selection methods.

\begin{figure}[!h]
   \centering
        \includegraphics[scale=.48]{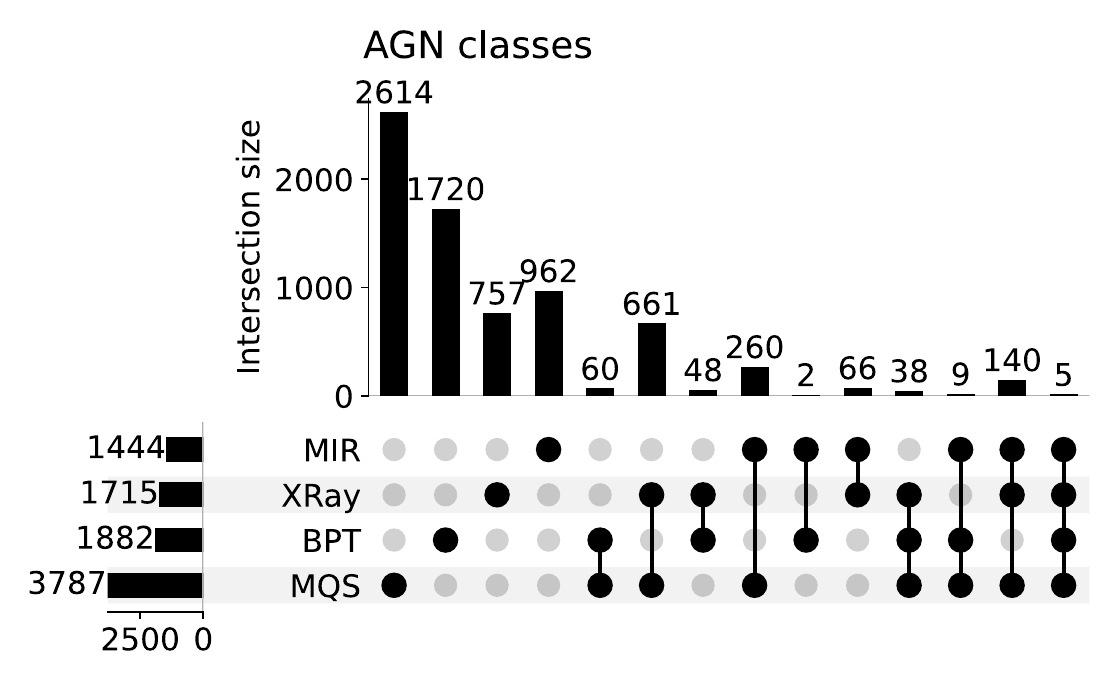}
    \caption{UpSet plot showing the intersection sizes among the four AGN classification methods used in this work: MIR-AGNs, BPT-AGNs, Xray-AGNs, and MQS-AGNs. The first four bars in the top represent number of galaxies with pure AGNs class; the rest of the top bars represent the number of galaxies belonging to a specific combination of these classes which are indicated by filled dots in the bottom matrix. The total number of galaxies identified by each individual method, are represented in the horizontal bars on the left. 
    }
    \label{UpSet}
  \end{figure}

Figure \ref{UpSet} presents the intersection sizes as bars, along with a matrix indicating the corresponding class combinations and the total number of galaxies in each sample, providing a comprehensive overview of both single and overlapping AGN identification methodologies. 

In the following, we refer to pure AGNs as galaxies identified by only one classification method, and to jointly classified AGNs as those identified by two or more methods simultaneously. The analysis shows that pure AGNs dominate the sample: only 17.6\% of the total catalogue of 7342 AGNs are classified by more than one method. Moreover, only 5 galaxies (0.07\% of the sample) are identified as AGNs by all four methods simultaneously.

Among the individual classes, X-ray AGNs exhibit the highest degree of overlap with other selections. Specifically, 958  galaxies --- corresponding to 55.8\% of the X-ray AGNs sample (1715 sources) --- are also identified by at least one additional method. This behaviour is consistent with a scenario in which X-ray AGNs represent intrinsically luminous or robust AGNs that are more likely to be independently detected through other diagnostics, such as optical or MIR.

 \begin{figure}[!h]
   \centering
        \includegraphics[scale=.45]{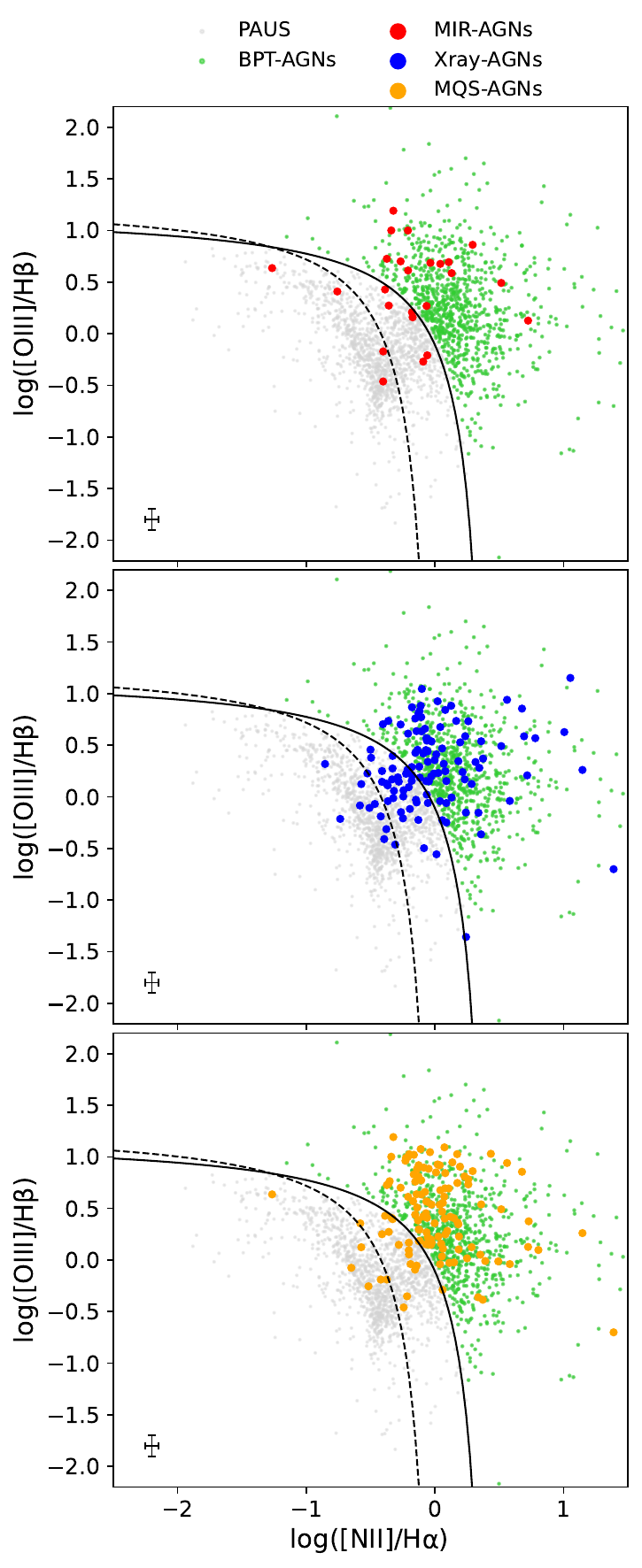}
    \caption{BPT diagram $\mathrm {log (NII6583}/\mathrm H_\alpha)$ vs. $\mathrm {log(OIII5006}/\mathrm H_\beta)$ for MIR-AGNs, BPT-AGNs, X-ray-AGNs, and MQS-AGNs. The mean error is presented in the lower left corner of the plot.  The solid line corresponds to the AGN demarcation from \citet{Kewley2001} and the dashed line the \citet{Kauffmann2003} line for pure star-forming galaxies.
    }
    \label{BPT}
  \end{figure}

As a complementary analysis, we studied the $\mathrm {log(NII6583}/\mathrm H_\alpha)$ vs. $\mathrm {log(OIII5006}/\mathrm H_\beta$ BPT diagram of 1882 BPT, 26 MIR, 146 Xray, and 141 MQS AGNs with suitable emission line measurements available in the \texttt{emissionLinesPort} table.

The results are shown in Fig. \ref{BPT}, where we also plotted 4246 PAUS galaxies with classification in the  \texttt{emissionLinesPort} table different from 'BLANK'. If we consider the \citet{Kewley2001} criteria to select AGNs, we found that 61\% MIR-AGNs, 65\% Xray-AGNs, 81\% MQS-AGNs are above the \citet{Kewley2001} demarcation line. Moreover if we consider the \citet{Kauffmann2003} demarcation line for composite galaxies, these fractions rise to 85\%, 86\% and 92\%, respectively.

These results indicate that a substantial fraction of AGNs selected through MIR, X-ray, and MQS criteria occupy regions of the BPT diagram traditionally associated with AGNs or composite ionization. While not implying a one-to-one correspondence between all classification methods, this comparison demonstrates a significant level of consistency between emission-line diagnostics and the multi-wavelength AGN selections adopted in this work.

\end{appendix}
\end{document}